\documentclass[ reprint, superscriptaddress, floatfix, amsmath,amssymb, aps, prx ]{revtex4-2}

\usepackage[utf8]{inputenc}
\usepackage{amsmath,amssymb,nccmath}
\usepackage{wrapfig}
\usepackage{graphicx}
\usepackage{bm}
\usepackage[colorlinks=true,linkcolor=blue,citecolor=blue,urlcolor=blue]{hyperref}
\usepackage{makecell}
\usepackage{siunitx}
\usepackage{cases}
\usepackage{tikz}
\usepackage{cancel}
\usetikzlibrary{backgrounds}

\newcommand{\citeasnoun}[1]{Ref.~\cite{#1}}

\newcommand{\Figref}[1]{Figure~\ref{fig:#1}}
\newcommand{\figref}[1]{Fig.~\ref{fig:#1}}
\renewcommand{\eqref}[1]{Eq.~(\ref{eq:#1})}
\newcommand{\Eqref}[1]{Equation~(\ref{eq:#1})}
\newcommand{\eqreftwo}[2]{Eqs.~(\ref{eq:#1},\ref{eq:#2})}
\newcommand{\eqrefrange}[2]{Eqs.~(\ref{eq:#1})--(\ref{eq:#2})}

\renewcommand{\Re}{\operatorname{Re}}
\renewcommand{\Im}{\operatorname{Im}}
\newcommand{\vect}[1]{\boldsymbol{\mathbf{#1}}}

\newcommand{\diag}{\operatorname{diag}}

\newcommand*{\psiinc}{\psi_{\rm inc}}

\newcommand*{\GO}{\Gamma_0}
\newcommand*{\DD}{\mathbb{D}}

\newcommand*{\Ev}{\vect{E}}
\newcommand*{\Hv}{\vect{H}}
\newcommand*{\Pv}{\vect{P}}
\newcommand*{\Mv}{\vect{M}}

\newcommand*{\St}{S_{\rm target}}

\newcommand*{\td}{\text{d}}
\newcommand*{\tomega}{\tilde{\omega}}
\newcommand*{\Hw}{H}
\newcommand*{\SM}{SM}

\newcommand{\hl}{}
\newcommand{\hll}[1]{{\color{blue}#1}}

\usepackage[english]{babel}
\makeatletter
\def\bbl@set@language#1{%
  \edef\languagename{%
    \ifnum\escapechar=\expandafter`\string#1\@empty
    \else\string#1\@empty\fi}%
  \@ifundefined{babel@language@alias@\languagename}{}{%
    \edef\languagename{\@nameuse{babel@language@alias@\languagename}}%
  }%
  \select@language{\languagename}%
  \expandafter\ifx\csname date\languagename\endcsname\relax\else
    \if@filesw
      \protected@write\@auxout{}{\string\select@language{\languagename}}%
      \bbl@for\bbl@tempa\BabelContentsFiles{%
        \addtocontents{\bbl@tempa}{\xstring\select@language{\languagename}}}%
      \bbl@usehooks{write}{}%
    \fi
  \fi}
\newcommand{\DeclareLanguageAlias}[2]{%
  \global\@namedef{babel@language@alias@#1}{#2}%
}
\makeatother

\DeclareLanguageAlias{en}{english}
\DeclareLanguageAlias{EN}{english}

\begin{document}

\preprint{APS/123-QED}

\title{Computational bounds to light--matter interactions via local conservation laws}

\author{Zeyu Kuang}
\affiliation{Department of Applied Physics and Energy Sciences Institute, Yale University, New Haven, Connecticut 06511, USA}
\author{Owen D. Miller}
\affiliation{Department of Applied Physics and Energy Sciences Institute, Yale University, New Haven, Connecticut 06511, USA}

\date{\today}

\begin{abstract}
    We develop a computational framework for identifying bounds to light--matter interactions, originating from polarization-current-based formulations of local conservation laws embedded in Maxwell's equations. We propose an iterative method for imposing only the maximally violated constraints, enabling rapid convergence to global bounds. Our framework can identify bounds to the minimum size of any scatterer that encodes a specific linear operator, given only its material properties, as we demonstrate for the optical computation of a discrete Fourier transform. It further resolves bounds on far-field scattering properties over any arbitrary bandwidth, where previous bounds diverge.
\end{abstract}

\maketitle

Nanoscale fabrication techniques, computational inverse design~\cite{Jensen2011,Miller2012a,Bendsoe2013,Molesky2018}, and fields from silicon photonics~\cite{Lalau-Keraly2013,Piggott2015,Yang2020,Sapra2020} to metasurface optics~\cite{yu2014flat,aieta2015multiwavelength,shrestha2018broadband,Zhan2019} are enabling transformative use of an unprecedented number of structural degrees of freedom in nanophotonics. An emerging critical need is an understanding of fundamental limits to what is possible, analogous to Shannon's bounds for digital communications~\cite{Shannon1949,Cover1999}. In this Letter, we identify an infinite set of local conservation laws that can form the foundation of a general framework for computational bounds to light--matter interactions. We show that this framework enables calculations of bounds for two pivotal applications, for which all previous approaches yield trivial (e.g., divergent) bounds. First, we identify computational bounds on the minimum size of a scatterer encoding any linear operator, demonstrated for an analog optical discrete Fourier transform (DFT). Second, we identify bounds on maximum far-field extinction over any bandwidth, resolving an important gap in power--bandwidth limits~\cite{Shim2019}. The local power-conservation laws identified here have immediate ramifications across nanophotonics; more generally, they appear to be extensible to linear partial differential equations across physics.

Bounds, or fundamental limits, identify what is possible in a complex design space. Beyond Shannon's bounds, well-known examples include the Carnot efficiency limit~\cite{Callen1985}, the Shockley--Queisser bounds in photovoltaics~\cite{Shockley1961}, the Bergman--Milton bounds in the theory of composites~\cite{Bergman1981,Milton1981,Milton2002}, and the Wheeler--Chu bounds on antenna quality factor~\cite{Wheeler1947,Chu1948}, among many more. In electromagnetism, for a long period of time there were very few bounds on general response functions (with a notable exception being sum rules on total response~\cite{gordon_1963, purcell_1969, mckellar_box_bohren_1982,sohl_gustafsson_kristensson_2007}), seemingly due to the complex and nonconvex nature of Maxwell's equations. Yet a flurry of recent results have suggested the possibility for general bounds~\cite{miller_johnson_rodriguez_2015,hugonin_besbes_ben-abdallah_2015,miller_polimeridis_reid_hsu_delacy_joannopoulos_soljacic_johnson_2016, miller_ilic_christensen_reid_atwater_joannopoulos_soljacic_johnson_2017,sanders_manjavacas_2018,yang_miller_christensen_joannopoulos_soljacic_2017,Yang2018, Shim2019, Michon2019, Zhang2019, Molesky2019a, ShimChung2020, Shim2020, Molesky2020a, Gustafsson2019, Molesky2020b, Kuang2020subm}, for quantities ranging from single-frequency scattering to radiation loss of free electrons, for bulk and 2D materials. Underlying all of these results is one or two energy-conservation laws, arising in various formulations of Maxwell's equations. Additional bounds have been identified via Lagrangian duality~\cite{Angeris2019,Trivedi2020} or physical approximations~\cite{Yu2010,Presutti2020,Chung2020}. Yet there are pivotal applications for which all of these approaches either do not apply or offer trivial bounds.

Here we identify an infinite set of conservation laws that must be satisfied by any solution of Maxwell's equations. These laws are ``domain-oblivious,'' i.e., once a designable region is specified, the constraints are valid for \emph{any} possible geometric structure in that region. Moreover, each conservation law is a quadratic form that is amenable to semidefinite relaxation~\cite{Laurent2005,Luo2010}. To accelerate the bound computations we develop an algorithm that automatically selects ideal constraints to impose. These bounds lack the intuition of analytical expressions, but they can provide significantly tighter limits.
\begin{figure*}[t]
    \includegraphics[width=0.8\textwidth]{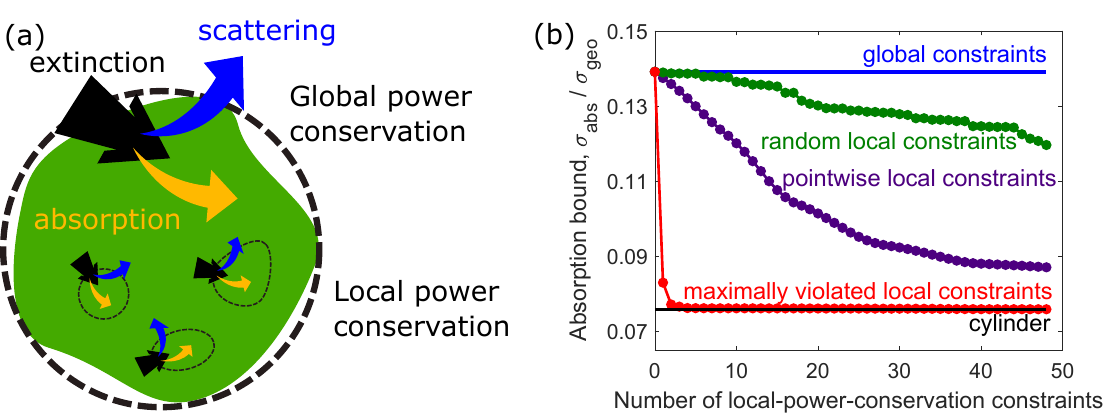}
    \centering
    \caption{(a) A freeform, homogeneous photonic scatterer within a designable region (outer dashed circle). Previous bounds utilized global conservation laws (large arrows). Here, we introduce a polarization-current-based formulation of local conservation laws that provide an infinite set of constraints for the identification of global bounds to light--matter interactions. (b) Example of local constraints (green, purple, red) tightening bounds from global constraints only (blue), for maximum absorption from a material with permittivity $\varepsilon = 12+0.1i$ in a region with diameter $d=0.18\lambda$. Our iterative method of selecting maximally violated constraints rapidly converges.}
    \label{fig:figure1}
\end{figure*}

\emph{Local Conservation Laws}---To start, we derive local conservation laws that must be satisfied by any Maxwell solution. These conservation laws manifest the complex Poynting theorem~\cite{Jackson1999} over any subdomain of a scatterer, but only when formulated in terms of induced polarization currents do they exhibit properties that enable global bounds. We consider any scattering problem comprising \hl{arbitrary sources and} arbitrary electric and/or magnetic material \hl{properties}. We use six-vector notation, concatenating electric and magnetic three-vectors for more concise expressions; for example, the electromagnetic fields $\psi$ and polarization currents $\phi$ are given by $\psi = \left(\begin{smallmatrix} \Ev \\ \Hv \end{smallmatrix}\right)$ and $\phi = \left(\begin{smallmatrix} \Pv \\ \Mv \end{smallmatrix}\right)$, and we use dimensionless units in which the speed of light is 1.

\hl{Physically, the conservation laws that form the foundation of our bounds arise from the complex Poynting theorem~\cite{Jackson1999}. As depicted in \figref{figure1}, Poynting's theorem must apply not only globally over an entire scatterer, but also locally at at any point within. We can rewrite the usual Poynting theorem (a function of the electromagnetic fields) in terms solely of the polarization currents $\phi$ induced across the scatterer. Ignoring reactive power momentarily, and considering only real power flow, the statement that at any point $x$ the local extinction must equal local absorption plus scattered power can be written:
    \begin{align}
        -\frac{\omega}{2} &\Im \left[\phi^\dagger(x)\psi_{\rm inc}(x)\right] = \frac{\omega}{2}\phi^\dagger(x)\Im \left[-\chi^{-1}\right]\phi(x) \label{eq:point_consv} \\ 
                           &+ \frac{\omega}{2}\Im \left[\phi^\dagger(x)\int_V \Gamma_0(x,x')\phi(x')\td x' \right] \nonumber
    \end{align}
    where the first term corresponds to extinction, the second term to absorption, and the last term to scattered power. \Eqref{point_consv} is the well-known optical theorem~\cite{Newton1976,Lytle2005,Jackson1999}, applied to an infinitesimal bounding sphere at point $x$. We can generalize this expression in three ways, physically argued here and rigorously justified in the {\SM}. First, we can allow for a complex-valued frequency and replace $\omega$ with its conjugate, \hl{$\omega^*$, which makes no difference at real frequencies but will be useful for bandwidth-averaged scattering below. Second, we can remove the imaginary part from \eqref{point_consv}, which then manifests the complex Poynting theorem, including reactive power conservation. Finally, instead of considering only a single position $x$, we can consider any linear combination of points as determined by taking the integral of \eqref{point_consv} against a weighting tensor $\DD(x)$ (which also isolates the polarization directions). Taken together, these generalizations comprise the constraints
\begin{align}
&-\frac{\omega^*}{2}\int_V \phi^\dagger(x) \DD(x) \psi_{\rm inc}(x)\td x \label{eq:vol_consv} \\ 
&= \frac{\omega^*}{2}\int_V \phi^\dagger(x) \DD(x) \td x \left[\int_V \Gamma_0(x,x')\phi(x')\td x' - \chi^{-1}\phi(x)\right].  \nonumber 
\end{align}}
Roughly speaking, \eqref{vol_consv} represents a linear combination of pointwise equalities from the complex Poynting theorem. To illuminate the algebraic structure of \eqref{vol_consv}, we can strip away the integrals and position dependencies by assuming any standard numerical discretization~\cite{Jin2011}, in which case $\phi$ and $\psi_{\rm inc}$ become vectors and $\DD$, $\Gamma_0$, and $\chi$ are matrices. After discretization, \eqref{vol_consv} is given in matrix notation by: 
\begin{align}
     \frac{\omega^*}{2} \phi^\dagger \DD\GO \phi - \frac{\omega^*}{2} \phi^\dagger \DD \chi^{-1} \phi = -\frac{\omega^*}{2} \phi^\dagger\DD\psi_{\rm inc} .
    \label{eq:optthm}
\end{align}
The complex-Poynting-theorem-based conservation laws of \eqreftwo{vol_consv}{optthm} satisfy two key properties that enable global bounds over all possible designs. First, they are ``domain-oblivious:'' within a designable region, \eqrefrange{point_consv}{optthm} must apply at every point regardless of whether it is part of the scattering domain or the background. Second, they are quadratic forms of the polarization currents, and therefore amenable to semidefinite programming, as we discuss below.}

\begin{figure*}[htb]
	\includegraphics[width=0.9\textwidth]{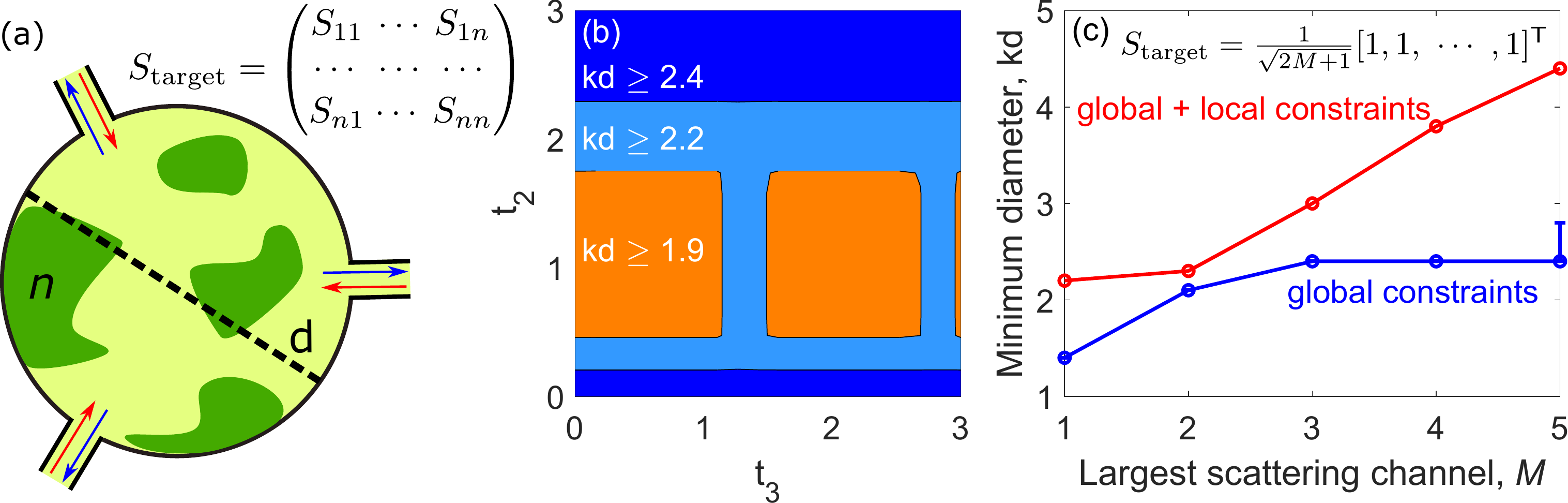}
    \centering
    \caption{(a) Photonic devices are often designed to achieve a specific ``target'' $S$ matrix in a compact form factor. Our bounds enable identification of the minimum diameter $d$ of any such device (relative to wavenumber $k$), for (b) nonuniform DFT-matrix implementation ($t_2$ and $t_3$ are parameters of \hl{the DFT matrix}) and (c) power splitters for a single input to $2M+1$ outgoing channels. In (b), \hl{each point in the image represents a unique DFT matrix, and} the colors indicate the \hl{minimum diameter for possibly achieving that scattering matrix.} In (c) it is evident that local constraints are required to identify feasible design regions as the required functionality increases in complexity. In (b,c) the channels are TE cylindrical waves and the material has refractive index $n=\sqrt{12}$.}
    \label{fig:figure2}
\end{figure*}

%Identifying a fundamental limit to the strongest possible response for a specific light--matter interaction requires maximization of a given metric subject to the constraints of Maxwell's equations, over the space of all possible designs. Such a specification is nonconvex and cannot be solved for a global optimum without simplification. Previous bounds tend to rely on extreme simplifications; the most general bounds, those of Refs.~\cite{Gustafsson2019,Molesky2020b,Kuang2020subm}, utilize one or both of the global versions of \eqref{vol_consv} in lieu of Maxwell's equations. Yet, as shown in the applications below, such constraints fail to produce useful bounds in important cases. If, however, one uses a large (or smart) selection of constraints from the infinite number of possible $\DD$ operators and constraints in \eqref{vol_consv}, significantly better bounds can be achieved.

\emph{Computational Bounds}---Any electromagnetic power- or momentum-flow objective function $f$ will be a linear or quadratic real-valued function of the polarization currents $\phi$, \hl{which in our matrix notation} can be written as $f(\phi) = \phi^\dagger \mathbb{A} \phi +  \Re\left(\beta^\dagger \phi\right)+c$, where $\mathbb{A}$ is any Hermitian \hl{matrix}, and $\beta$ and $c$ are any vector and constant, respectively. \hl{To identify bounds for any objective $f$, we replace the Maxwell-equation constraint (which is not domain-oblivious) with a finite number of constraints of the form of \eqref{optthm}, each with a unique $\DD$ matrix given by $\DD_j$ for the $j^{\rm th}$ constraint. Then, a bound on the maximum achievable $f$ is given by the solution of:}
\begin{equation}
    \begin{aligned}
        & \underset{\phi}{\text{max.}}
        & & f(\phi) = \phi^\dagger \mathbb{A} \phi +  \Re\left(\beta^\dagger \phi\right)+c \\
        & \text{s.t.}
        & & \phi^\dagger \Re\left\{\DD_j\omega^*(\Gamma_0-\chi^{-1})\right\} \phi = -\Re \left(\omega^*\phi^\dagger\DD_j\psiinc\right).
    \end{aligned}
    \label{eq:opt_prob}
\end{equation}
\hl{The domain of the optimization variable $\phi$ is the space of coefficients of the basis functions extending over an entire designable region (e.g. the dashed-circled cylinder in \figref{figure1} (a))}. We have taken only the real part of \eqref{vol_consv} because $\DD \rightarrow i\DD$ accounts for the imaginary part. \Eqref{opt_prob} is a key result: \hl{it is a formulation of maximum response, subject to all possible local-power conservation laws, as} a quadratically constraint quadratic program, i.e. a QCQP optimization problem~\cite{Boyd2004,Luo2010}. By virtue of the domain-oblivious property of the constraints, it applies to all possible designs within the designable domain. A bound on the solution of \eqref{opt_prob} can be found by standard techniques that relax the original, quadratic program to a higher-dimensional linear program over semidefinite matrices, i.e., a semidefinite program (SDP)~\cite{Laurent2005,Luo2010}, which can be solved by interior-point methods~\cite{Vandenberghe1996,Boyd2004}. Such transformations of QCQP's have led to meaningful bounds in many areas of engineering~\cite{Goemans1995,Vandenberghe1996,Tan2001,Biswas2006,Luo2010,Gershman2010}; we leave the details of the transformation of \eqref{opt_prob} to the {\SM}. The final solution represents a global, unsurpassable bound for any electromagnetic scattering response.

It is computationally prohibitive to impose the infinitely many constraints of \eqref{vol_consv}. We propose an iterative \hl{algorithm for identifying which subset of constraints to use.} One should start with the two $\DD$ matrices that correspond to global power conservation, i.e. the identity tensor and the identity tensor multiplied by $i$, which correspond to \hl{reactive- and real-power} conservation, respectively. (The latter leads to a positive semidefinite quadratic form and is crucial to restricting the magnitude of the solutions.) \hl{As a first iteration, we use only those two} $\DD$ matrices \hl{to find an initial bound for} \eqref{opt_prob}, as well as the \hl{first-iteration} optimal polarization currents, $\phi_{\rm opt,1}$. From those currents, we can identify out of all possible remaining $\DD$-matrix constraints, which ones are ``most violated'' by $\phi_{\rm opt,1}$, i.e. \hl{which constraint is farthest from zero} (under the $L_2$ norm). \hl{The constraint for the corresponding $\DD$ matrix is then added to the constraint set, and a second iteration is run, identifying new bounds and new optimal polarization currents. This process proceeds iteratively until convergence. Straightforward linear algebra shows (cf. {\SM}) that after iteration $j$, with optimal currents $\phi_j$, the next constraint to add is the one with $\DD$ matrix
\begin{align}
    \DD_{j+1} = \omega \diag\left[\phi_{\rm opt,j} \phi_{\rm opt,j}^\dagger \left(\Gamma_0-\chi^{-1}\right)^\dagger + \phi_{\rm opt,j}\psiinc \right],
\end{align}}
where ``$\diag$'' is the diagonal (in space) matrix comprising the diagonal elements of its matrix argument. \Figref{figure1}(b) demonstrates the utility of this method \hl{of maximally violated constraints} for bounding the TE absorption cross-section $\sigma_{\rm abs}$ of a dielectric scatterer \hl{of any shape} occupying a wavelength-scale cylindrical design region. \hl{The designable region} \hl{need not be symmetric; in the {\SM} we include an example with a triangular region.} Whereas the global constraints (blue) are significantly larger than the response of a \hl{cylindrical scatterer} (black), including local constraints shows that one can clearly identify tighter bounds. Yet both randomly chosen $\DD$ matrices (green) and \hl{spatially pointwise}, delta-function-based $\DD$ matrices (purple) show slow convergence. 
The iterative method via maximally violated constraints shows rapid convergence, requiring only two local constraints.
\hl{The spatial patterns of both the optimal current distribution and local constraints are shown in the {\SM}.}
From this method we can clearly identify the cylinder as a globally optimal structure for that material and design region.

\emph{S-Matrix Feasibility}---To demonstrate the power of this framework, we consider a fundamental question in the fields of analog optical computing~\cite{Silva2014,Pors2015,Zhu2017,Kwon2018,Estakhri2019} and metasurfaces~\cite{yu2014flat,aieta2015multiwavelength,shrestha2018broadband}: what is the minimum size of a scatterer that achieves a desired scattering matrix $\St$? A generic setup is depicted in \figref{figure2}(a). The target $S$ matrix could manifest lens focusing or meta-optical computing, for example. The objective, then, is to minimize the relative difference between the achievable and target $S$ matrices, i.e., $f_{\rm obj} = \left\| S - \St \right\|^2 / \left\|\St\right\|^2$, where $\|\cdot\|$ denote\hll{s the} Frobenius norm. It is straightforward to write this objective in the form appearing in \eqref{opt_prob}, as the $S$ matrix elements are linear in the polarization currents and the objective is a quadratic form (cf. {\SM}). Then, to determine the minimum feasible size for implementing $\St$, we can compute the bound on the smallest error between $S$ and $\St$, and define an acceptable-accuracy threshold (1\%) below which the device \hl{exhibits the desired functionality with sufficient fidelity.}

We apply our framework to two such problems, both of which comprise two-dimensional, nonmagnetic scatterers with refractive index $n=\sqrt{12}$, discretized by the discrete dipole approximation (DDA)~\cite{Purcell1973,Draine1994}. In the first, we identify the smallest domain within which a scatterer can possibly act as a discrete Fourier transform (DFT) operator over three TE cylindrical-wave channels (\hl{cf.} {\SM}). \hl{The DFT is the foundation for discrete Fourier analysis} \hl{and many other practical applications~\cite{strang_wavelets_1994}. } \hl{With uniform frequencies and nonuniform sample points $t_1$, $t_2$, and $t_3$ (and $t_1$ is fixed as a reference to be $t_1 = 0$), a target $S$ matrix that acts as a DFT can be represented as}\hll{~\cite{rao_nonuniform_2010}}:
\begin{equation}
    S_{\rm target}(t_2,t_3) = \frac{1}{\sqrt{3}}
    \begin{pmatrix}
        1 & 1 & 1\\
        1 & e^{-2\pi it_2/3} & e^{-2\pi it_3/3}\\
        1 & e^{-4\pi it_2/3} & e^{-4\pi it_3/3}
    \end{pmatrix}.
    \label{eq:S_target1}
\end{equation}
\hl{\Figref{figure2}(b) shows the bound-based feasibility map for implementing such an $S$ matrix. Each point in the grid represents a unique DFT matrix (prescribed by the values of $t_2$ and $t_3$), and the color indicates the smallest diameter $d$, relative to wavenumber $k$, of a structure that can possibly exhibit the desired DFT-based scattering matrix (at 99\% fidelity). There is no structure, with any type of patterning, that can act as a DFT matrix if its diameter is smaller than that specified in \figref{figure2}(b). Thus our approach enables bounds on the minimal possible size of an optical element implementing specific functionality.} A related calculation is shown in \figref{figure2}(c). In that case we consider a target $S$ matrix \hll{for} a power splitter, directing a single incident wave equally into outgoing spherical-wave channels index by $m$ (where $m$ is the angular index, $m=-M,\ldots,M$). We depict the minimum diameter as a function of the number of scattering channels, both for the global-constraint-only approach (blue) and our new approach with local constraints. (The error bar indicates a numerical instability in the global-constraint-only approach, cf. {\SM}.) \hl{Whereas the global-constraint-only approach unphysically converges to wavelength scale as the number of channels increases}, our new approach \hl{predicts an unavoidable increase in the diameter of the power splitter, representing the first such capability for capturing minimum-size increases with increasing complexity.}

\begin{figure}[htb]
	\includegraphics[width=0.5\textwidth]{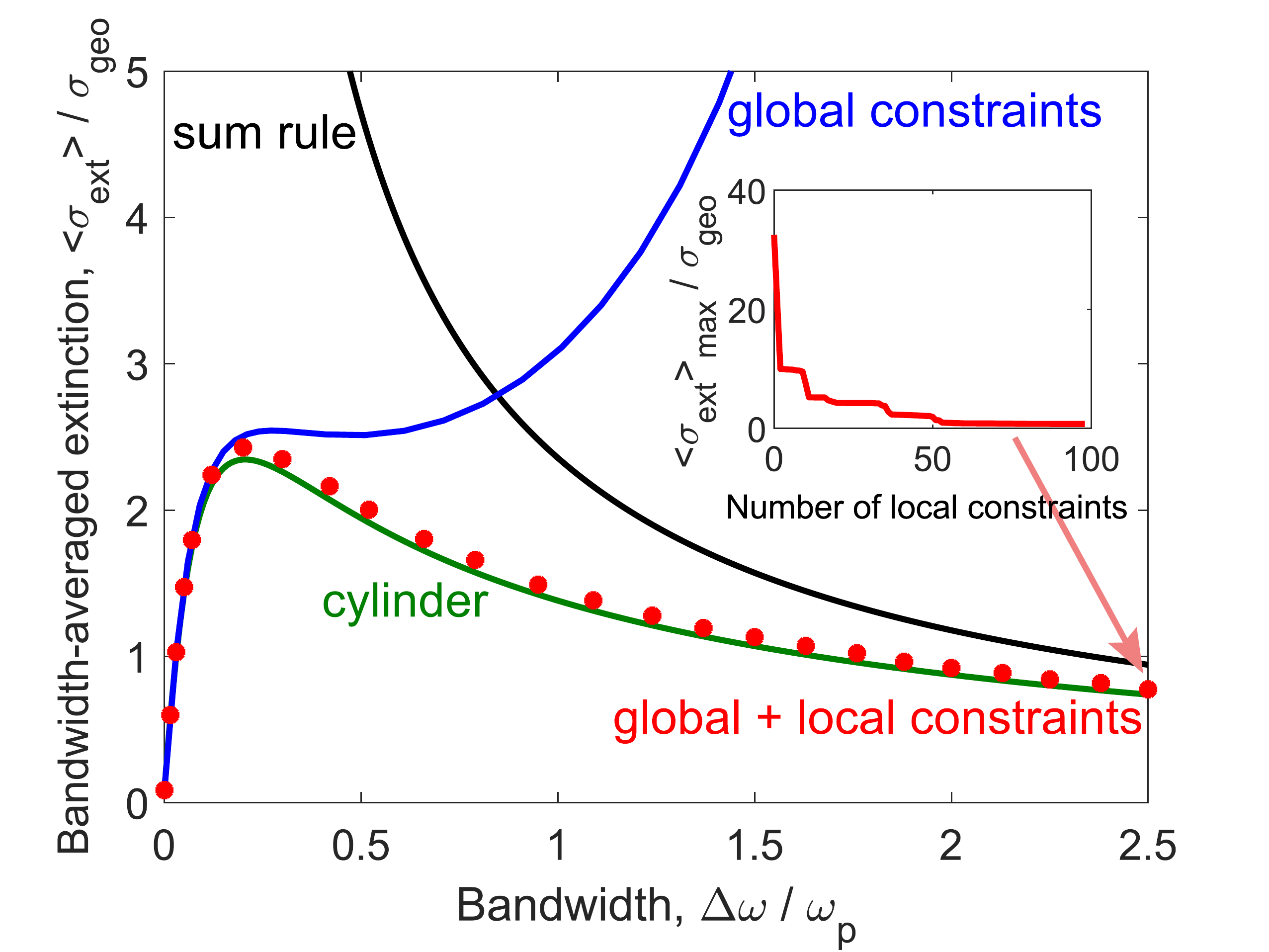}
    \centering
    \caption{Bounds on maximal bandwidth-averaged extinction, $\langle \sigma_{\rm ext}\rangle_{\rm max}$, as a function of bandwidth $\Delta\omega$ for a lossless Lorentz--Drude material with plasma frequency $\omega_p$ and oscillator frequency $\omega_c = 0.3015\omega_p$, which is chosen such that the permittivity is 12 at a center frequency $\omega_0 = 0.05\omega_p$. The bounds are normalized to the geometric cross-section $\sigma_{\rm geo}$ of the designable region, a cylinder with diameter $d=3/\omega_p$. While known sum rules (black) and global-constraint bounds (blue) are loose for many bandwidths, utilizing local constraints (convergence shown in inset) enables apparently tight bounds across all bandwidths.}
    \label{fig:figure3}
\end{figure}

\emph{Far-field power--bandwidth limits}---The local-constraint bound framework resolves another outstanding question: how large can far-field scattering be over an arbitrary bandwidth $\Delta \omega$? In \citeasnoun{Shim2019}, bounds for near-field average-bandwidth response were derived using global constraints at a complex frequency; yet it was noted that the same technique fails in the far field (it exhibits an unphysical divergence). A feature of \eqref{vol_consv} is that the local conservation laws can also be applied at complex frequencies, as the inclusion of \hl{the conjugate frequency $\omega^*$} leads to operators that are positive semidefinite over the whole upper half of the complex-frequency plane, by passivity (cf. {\SM}). %({\SM}).

A prototypical example to consider is the maximum extinction cross-section, $\sigma_{\rm ext}(\omega)$, from a given material over a bandwidth $\Delta\omega$. Using contour-integral techniques from Refs.~(\cite{Hashemi2012,Shim2019}), the average extinction around a center frequency $\omega_0$, over a bandwidth $\Delta\omega$, as measured by integration against a Lorentzian window function, $\hl{\Hw}(\omega)=\frac{\Delta\omega/\pi}{(\omega-\omega_0)^2+\Delta\omega^2}$, can be written as the evaluation of a single scattering amplitude at a \emph{complex} frequency $\tomega$ (cf. {\SM}):
\begin{align}
    \langle \sigma_{\rm ext}\rangle  &= \int_{-\infty}^{+\infty} \sigma_{\rm ext}(\omega)\hl{\Hw}(\omega)\td\omega \nonumber \\
	 &=\Im\left[\tomega \psi^T_{\rm inc}(-\tomega)\phi(\tomega)\right], \label{eq:complex_ext}
\end{align}
where $\tomega = \omega_0 + i\Delta\omega$. \Eqref{complex_ext} is a linear objective function of the form required by \eqref{opt_prob}, evaluated at a complex frequency. By imposing the global- and local-conservation constraints at the complex frequency $\tomega$, we can identify bounds to the bandwidth-averaged far-field response. \Figref{figure3} shows the results of such a computation for a lossless Lorentz--Drude material \hl{(with plasma frequency $\omega_p$)} in a designable region with diameter $d = 3/\omega_p$. Included in the figure is a bound on average extinction from a known all-frequency sum rule~\cite{gordon_1963,Yang2015} (black), which \hl{is} descriptive in the infinite-bandwidth limit, and the global-constraint-only bounds (blue), which are useful in the small-bandwidth limit, but \hl{each diverges} in the opposite limits. Through the use of global and local constraints (red), we can identify bounds over any bandwidth of interest, and we find that a cylindrical scatterer is nearly globally optimal.

\emph{Conclusions}---We have shown that local conservation laws enable computational bounds to light--matter interactions. The demonstrated bounds for optical analog computing and power--bandwidth limits are suggestive of a wide array of future possible applications. From the perspective of identifying feasible design volumes for target scattering matrices, a natural extension is to large-area, broadband metalenses. It is clear that there are tradeoffs between diameter, bandwidth, and efficiency, but the optimal architecture and form factor is unknown. Our bounds may resolve the Pareto frontier. Similarly, the power--bandwidth limits have natural applications in photovoltaics~\cite{Yu2010,Ganapati2014,Buddhiraju2017,Benzaouia2019} and ultrafast optics~\cite{Weiner2011,Agrawal2013,Wang2008}. 

\hl{There are two key areas for improvement looking forward: the first is to nonlinear optics and nonlinear physics, where conservation laws analogous to \eqref{optthm} would not have the quadratic structure that enabled semidefinite-programming-based bounds here. The second is to identify faster computational schemes, such as those used in ``fast solvers''~\cite{rokhlin_rapid_1985, harrington_field_1993-1, johnson_block-iterative_2001}, as the computational cost of semidefinite relaxations prohibited our exploration of structures far larger than wavelength scale. Overcoming both of these limitations would open interesting possibilities for applications ranging from quantum dynamics to large-scale meta-optics.}

\emph{Acknowledgments}---This work was supported by the Army Research Office under grant number W911NF-19-1-0279.

\bibliography{lc_bib}

\begin{thebibliography}{79}%
\makeatletter
\providecommand \@ifxundefined [1]{%
 \@ifx{#1\undefined}
}%
\providecommand \@ifnum [1]{%
 \ifnum #1\expandafter \@firstoftwo
 \else \expandafter \@secondoftwo
 \fi
}%
\providecommand \@ifx [1]{%
 \ifx #1\expandafter \@firstoftwo
 \else \expandafter \@secondoftwo
 \fi
}%
\providecommand \natexlab [1]{#1}%
\providecommand \enquote  [1]{``#1''}%
\providecommand \bibnamefont  [1]{#1}%
\providecommand \bibfnamefont [1]{#1}%
\providecommand \citenamefont [1]{#1}%
\providecommand \href@noop [0]{\@secondoftwo}%
\providecommand \href [0]{\begingroup \@sanitize@url \@href}%
\providecommand \@href[1]{\@@startlink{#1}\@@href}%
\providecommand \@@href[1]{\endgroup#1\@@endlink}%
\providecommand \@sanitize@url [0]{\catcode `\\12\catcode `\$12\catcode
  `\&12\catcode `\#12\catcode `\^12\catcode `\_12\catcode `\%12\relax}%
\providecommand \@@startlink[1]{}%
\providecommand \@@endlink[0]{}%
\providecommand \url  [0]{\begingroup\@sanitize@url \@url }%
\providecommand \@url [1]{\endgroup\@href {#1}{\urlprefix }}%
\providecommand \urlprefix  [0]{URL }%
\providecommand \Eprint [0]{\href }%
\providecommand \doibase [0]{https://doi.org/}%
\providecommand \selectlanguage [0]{\@gobble}%
\providecommand \bibinfo  [0]{\@secondoftwo}%
\providecommand \bibfield  [0]{\@secondoftwo}%
\providecommand \translation [1]{[#1]}%
\providecommand \BibitemOpen [0]{}%
\providecommand \bibitemStop [0]{}%
\providecommand \bibitemNoStop [0]{.\EOS\space}%
\providecommand \EOS [0]{\spacefactor3000\relax}%
\providecommand \BibitemShut  [1]{\csname bibitem#1\endcsname}%
\let\auto@bib@innerbib\@empty
%</preamble>
\bibitem [{\citenamefont {Jensen}\ and\ \citenamefont
  {Sigmund}(2011)}]{Jensen2011}%
  \BibitemOpen
  \bibfield  {author} {\bibinfo {author} {\bibfnamefont {J.~S.}\ \bibnamefont
  {Jensen}}\ and\ \bibinfo {author} {\bibfnamefont {O.}~\bibnamefont
  {Sigmund}},\ }\bibfield  {title} {\bibinfo {title} {{Topology optimization
  for nano-photonics}},\ }\href {https://doi.org/10.1002/lpor.201000014}
  {\bibfield  {journal} {\bibinfo  {journal} {Laser {\&} Photonics Rev.}\
  }\textbf {\bibinfo {volume} {5}},\ \bibinfo {pages} {308} (\bibinfo {year}
  {2011})}\BibitemShut {NoStop}%
\bibitem [{\citenamefont {Miller}(2012)}]{Miller2012a}%
  \BibitemOpen
  \bibfield  {author} {\bibinfo {author} {\bibfnamefont {O.~D.}\ \bibnamefont
  {Miller}},\ }\emph {\bibinfo {title} {{Photonic Design: From Fundamental
  Solar Cell Physics to Computational Inverse Design}}},\ \href
  {http://arxiv.org/abs/1308.0212} {Ph.D. thesis},\ \bibinfo  {school}
  {University of California, Berkeley} (\bibinfo {year} {2012})\BibitemShut
  {NoStop}%
\bibitem [{\citenamefont {Bendsoe}\ and\ \citenamefont
  {Sigmund}(2013)}]{Bendsoe2013}%
  \BibitemOpen
  \bibfield  {author} {\bibinfo {author} {\bibfnamefont {M.~P.}\ \bibnamefont
  {Bendsoe}}\ and\ \bibinfo {author} {\bibfnamefont {O.}~\bibnamefont
  {Sigmund}},\ }\href@noop {} {\emph {\bibinfo {title} {{Topology optimization:
  theory, methods, and applications}}}}\ (\bibinfo  {publisher} {Springer
  Science {\&} Business Media},\ \bibinfo {year} {2013})\BibitemShut {NoStop}%
\bibitem [{\citenamefont {Molesky}\ \emph {et~al.}(2018)\citenamefont
  {Molesky}, \citenamefont {Lin}, \citenamefont {Piggott}, \citenamefont {Jin},
  \citenamefont {Vuckovi{\'{c}}},\ and\ \citenamefont
  {Rodriguez}}]{Molesky2018}%
  \BibitemOpen
  \bibfield  {author} {\bibinfo {author} {\bibfnamefont {S.}~\bibnamefont
  {Molesky}}, \bibinfo {author} {\bibfnamefont {Z.}~\bibnamefont {Lin}},
  \bibinfo {author} {\bibfnamefont {A.~Y.}\ \bibnamefont {Piggott}}, \bibinfo
  {author} {\bibfnamefont {W.}~\bibnamefont {Jin}}, \bibinfo {author}
  {\bibfnamefont {J.}~\bibnamefont {Vuckovi{\'{c}}}},\ and\ \bibinfo {author}
  {\bibfnamefont {A.~W.}\ \bibnamefont {Rodriguez}},\ }\bibfield  {title}
  {\bibinfo {title} {{Inverse design in nanophotonics}},\ }\href
  {https://doi.org/10.1038/s41566-018-0246-9} {\bibfield  {journal} {\bibinfo
  {journal} {Nature Photonics}\ }\textbf {\bibinfo {volume} {12}},\ \bibinfo
  {pages} {659} (\bibinfo {year} {2018})},\ \Eprint
  {https://arxiv.org/abs/1801.06715} {1801.06715} \BibitemShut {NoStop}%
\bibitem [{\citenamefont {Lalau-Keraly}\ \emph {et~al.}(2013)\citenamefont
  {Lalau-Keraly}, \citenamefont {Bhargava}, \citenamefont {Miller},\ and\
  \citenamefont {Yablonovitch}}]{Lalau-Keraly2013}%
  \BibitemOpen
  \bibfield  {author} {\bibinfo {author} {\bibfnamefont {C.~M.}\ \bibnamefont
  {Lalau-Keraly}}, \bibinfo {author} {\bibfnamefont {S.}~\bibnamefont
  {Bhargava}}, \bibinfo {author} {\bibfnamefont {O.~D.}\ \bibnamefont
  {Miller}},\ and\ \bibinfo {author} {\bibfnamefont {E.}~\bibnamefont
  {Yablonovitch}},\ }\bibfield  {title} {\bibinfo {title} {{Adjoint shape
  optimization applied to electromagnetic design}},\ }\href
  {https://doi.org/10.1364/OE.21.021693} {\bibfield  {journal} {\bibinfo
  {journal} {Opt. Express}\ }\textbf {\bibinfo {volume} {21}},\ \bibinfo
  {pages} {21693} (\bibinfo {year} {2013})}\BibitemShut {NoStop}%
\bibitem [{\citenamefont {Piggott}\ \emph {et~al.}(2015)\citenamefont
  {Piggott}, \citenamefont {Lu}, \citenamefont {Lagoudakis}, \citenamefont
  {Petykiewicz}, \citenamefont {Babinec},\ and\ \citenamefont
  {Vuckovi{\'{c}}}}]{Piggott2015}%
  \BibitemOpen
  \bibfield  {author} {\bibinfo {author} {\bibfnamefont {A.~Y.}\ \bibnamefont
  {Piggott}}, \bibinfo {author} {\bibfnamefont {J.}~\bibnamefont {Lu}},
  \bibinfo {author} {\bibfnamefont {K.~G.}\ \bibnamefont {Lagoudakis}},
  \bibinfo {author} {\bibfnamefont {J.}~\bibnamefont {Petykiewicz}}, \bibinfo
  {author} {\bibfnamefont {T.~M.}\ \bibnamefont {Babinec}},\ and\ \bibinfo
  {author} {\bibfnamefont {J.}~\bibnamefont {Vuckovi{\'{c}}}},\ }\bibfield
  {title} {\bibinfo {title} {{Inverse design and demonstration of a compact and
  broadband on-chip wavelength demultiplexer}},\ }\href
  {https://doi.org/10.1038/nphoton.2015.69} {\bibfield  {journal} {\bibinfo
  {journal} {Nat. Photonics}\ }\textbf {\bibinfo {volume} {9}},\ \bibinfo
  {pages} {374} (\bibinfo {year} {2015})},\ \Eprint
  {https://arxiv.org/abs/1504.00095} {1504.00095} \BibitemShut {NoStop}%
\bibitem [{\citenamefont {Yang}\ \emph {et~al.}(2020)\citenamefont {Yang},
  \citenamefont {Skarda}, \citenamefont {Cotrufo}, \citenamefont {Dutt},
  \citenamefont {Ahn}, \citenamefont {Sawaby}, \citenamefont {Vercruysse},
  \citenamefont {Arbabian}, \citenamefont {Fan}, \citenamefont {Al{\`{u}}},\
  and\ \citenamefont {Vu{\v{c}}kovi{\'{c}}}}]{Yang2020}%
  \BibitemOpen
  \bibfield  {author} {\bibinfo {author} {\bibfnamefont {K.~Y.}\ \bibnamefont
  {Yang}}, \bibinfo {author} {\bibfnamefont {J.}~\bibnamefont {Skarda}},
  \bibinfo {author} {\bibfnamefont {M.}~\bibnamefont {Cotrufo}}, \bibinfo
  {author} {\bibfnamefont {A.}~\bibnamefont {Dutt}}, \bibinfo {author}
  {\bibfnamefont {G.~H.}\ \bibnamefont {Ahn}}, \bibinfo {author} {\bibfnamefont
  {M.}~\bibnamefont {Sawaby}}, \bibinfo {author} {\bibfnamefont
  {D.}~\bibnamefont {Vercruysse}}, \bibinfo {author} {\bibfnamefont
  {A.}~\bibnamefont {Arbabian}}, \bibinfo {author} {\bibfnamefont
  {S.}~\bibnamefont {Fan}}, \bibinfo {author} {\bibfnamefont {A.}~\bibnamefont
  {Al{\`{u}}}},\ and\ \bibinfo {author} {\bibfnamefont {J.}~\bibnamefont
  {Vu{\v{c}}kovi{\'{c}}}},\ }\bibfield  {title} {\bibinfo {title}
  {{Inverse-designed non-reciprocal pulse router for chip-based LiDAR}},\
  }\href {https://doi.org/10.1038/s41566-020-0606-0} {\bibfield  {journal}
  {\bibinfo  {journal} {Nat. Photonics}\ }\textbf {\bibinfo {volume} {14}},\
  \bibinfo {pages} {369} (\bibinfo {year} {2020})}\BibitemShut {NoStop}%
\bibitem [{\citenamefont {Sapra}\ \emph {et~al.}(2020)\citenamefont {Sapra},
  \citenamefont {Yang}, \citenamefont {Vercruysse}, \citenamefont {Leedle},
  \citenamefont {Black}, \citenamefont {England}, \citenamefont {Su},
  \citenamefont {Trivedi}, \citenamefont {Miao}, \citenamefont {Solgaard},
  \citenamefont {Byer},\ and\ \citenamefont
  {Vu{\v{c}}kovi{\'{c}}}}]{Sapra2020}%
  \BibitemOpen
  \bibfield  {author} {\bibinfo {author} {\bibfnamefont {N.~V.}\ \bibnamefont
  {Sapra}}, \bibinfo {author} {\bibfnamefont {K.~Y.}\ \bibnamefont {Yang}},
  \bibinfo {author} {\bibfnamefont {D.}~\bibnamefont {Vercruysse}}, \bibinfo
  {author} {\bibfnamefont {K.~J.}\ \bibnamefont {Leedle}}, \bibinfo {author}
  {\bibfnamefont {D.~S.}\ \bibnamefont {Black}}, \bibinfo {author}
  {\bibfnamefont {R.~J.}\ \bibnamefont {England}}, \bibinfo {author}
  {\bibfnamefont {L.}~\bibnamefont {Su}}, \bibinfo {author} {\bibfnamefont
  {R.}~\bibnamefont {Trivedi}}, \bibinfo {author} {\bibfnamefont
  {Y.}~\bibnamefont {Miao}}, \bibinfo {author} {\bibfnamefont {O.}~\bibnamefont
  {Solgaard}}, \bibinfo {author} {\bibfnamefont {R.~L.}\ \bibnamefont {Byer}},\
  and\ \bibinfo {author} {\bibfnamefont {J.}~\bibnamefont
  {Vu{\v{c}}kovi{\'{c}}}},\ }\bibfield  {title} {\bibinfo {title} {{On-chip
  integrated laser-driven particle accelerator}},\ }\href
  {https://doi.org/10.1126/science.aay5734} {\bibfield  {journal} {\bibinfo
  {journal} {Science}\ }\textbf {\bibinfo {volume} {367}},\ \bibinfo {pages}
  {79} (\bibinfo {year} {2020})}\BibitemShut {NoStop}%
\bibitem [{\citenamefont {Yu}\ and\ \citenamefont
  {Capasso}(2014)}]{yu2014flat}%
  \BibitemOpen
  \bibfield  {author} {\bibinfo {author} {\bibfnamefont {N.}~\bibnamefont
  {Yu}}\ and\ \bibinfo {author} {\bibfnamefont {F.}~\bibnamefont {Capasso}},\
  }\bibfield  {title} {\bibinfo {title} {Flat optics with designer
  metasurfaces},\ }\href {https://doi.org/10.1038/nmat3839} {\bibfield
  {journal} {\bibinfo  {journal} {Nature materials}\ }\textbf {\bibinfo
  {volume} {13}},\ \bibinfo {pages} {139} (\bibinfo {year} {2014})}\BibitemShut
  {NoStop}%
\bibitem [{\citenamefont {Aieta}\ \emph {et~al.}(2015)\citenamefont {Aieta},
  \citenamefont {Kats}, \citenamefont {Genevet},\ and\ \citenamefont
  {Capasso}}]{aieta2015multiwavelength}%
  \BibitemOpen
  \bibfield  {author} {\bibinfo {author} {\bibfnamefont {F.}~\bibnamefont
  {Aieta}}, \bibinfo {author} {\bibfnamefont {M.~A.}\ \bibnamefont {Kats}},
  \bibinfo {author} {\bibfnamefont {P.}~\bibnamefont {Genevet}},\ and\ \bibinfo
  {author} {\bibfnamefont {F.}~\bibnamefont {Capasso}},\ }\bibfield  {title}
  {\bibinfo {title} {Multiwavelength achromatic metasurfaces by dispersive
  phase compensation},\ }\href {https://doi.org/10.1126/science.aaa2494}
  {\bibfield  {journal} {\bibinfo  {journal} {Science}\ }\textbf {\bibinfo
  {volume} {347}},\ \bibinfo {pages} {1342} (\bibinfo {year}
  {2015})}\BibitemShut {NoStop}%
\bibitem [{\citenamefont {Shrestha}\ \emph {et~al.}(2018)\citenamefont
  {Shrestha}, \citenamefont {Overvig}, \citenamefont {Lu}, \citenamefont
  {Stein},\ and\ \citenamefont {Yu}}]{shrestha2018broadband}%
  \BibitemOpen
  \bibfield  {author} {\bibinfo {author} {\bibfnamefont {S.}~\bibnamefont
  {Shrestha}}, \bibinfo {author} {\bibfnamefont {A.~C.}\ \bibnamefont
  {Overvig}}, \bibinfo {author} {\bibfnamefont {M.}~\bibnamefont {Lu}},
  \bibinfo {author} {\bibfnamefont {A.}~\bibnamefont {Stein}},\ and\ \bibinfo
  {author} {\bibfnamefont {N.}~\bibnamefont {Yu}},\ }\bibfield  {title}
  {\bibinfo {title} {Broadband achromatic dielectric metalenses},\ }\href
  {https://doi.org/10.1038/s41377-018-0078-x} {\bibfield  {journal} {\bibinfo
  {journal} {Light: Science \& Applications}\ }\textbf {\bibinfo {volume}
  {7}},\ \bibinfo {pages} {85} (\bibinfo {year} {2018})}\BibitemShut {NoStop}%
\bibitem [{\citenamefont {Zhan}\ \emph {et~al.}(2019)\citenamefont {Zhan},
  \citenamefont {Gibson}, \citenamefont {Whitehead}, \citenamefont {Smith},
  \citenamefont {Hendrickson},\ and\ \citenamefont {Majumdar}}]{Zhan2019}%
  \BibitemOpen
  \bibfield  {author} {\bibinfo {author} {\bibfnamefont {A.}~\bibnamefont
  {Zhan}}, \bibinfo {author} {\bibfnamefont {R.}~\bibnamefont {Gibson}},
  \bibinfo {author} {\bibfnamefont {J.}~\bibnamefont {Whitehead}}, \bibinfo
  {author} {\bibfnamefont {E.}~\bibnamefont {Smith}}, \bibinfo {author}
  {\bibfnamefont {J.~R.}\ \bibnamefont {Hendrickson}},\ and\ \bibinfo {author}
  {\bibfnamefont {A.}~\bibnamefont {Majumdar}},\ }\bibfield  {title} {\bibinfo
  {title} {{Controlling three-dimensional optical fields via inverse Mie
  scattering}},\ }\href {https://doi.org/10.1126/sciadv.aax4769} {\bibfield
  {journal} {\bibinfo  {journal} {Sci. Adv.}\ }\textbf {\bibinfo {volume}
  {5}},\ \bibinfo {pages} {1} (\bibinfo {year} {2019})}\BibitemShut {NoStop}%
\bibitem [{\citenamefont {Shannon}\ and\ \citenamefont
  {Weaver}(1949)}]{Shannon1949}%
  \BibitemOpen
  \bibfield  {author} {\bibinfo {author} {\bibfnamefont {C.~E.}\ \bibnamefont
  {Shannon}}\ and\ \bibinfo {author} {\bibfnamefont {W.}~\bibnamefont
  {Weaver}},\ }\href@noop {} {\emph {\bibinfo {title} {{The Mathematical Theory
  of Communication}}}}\ (\bibinfo  {publisher} {Univ. of Illinois Press},\
  \bibinfo {address} {Urbana, IL},\ \bibinfo {year} {1949})\BibitemShut
  {NoStop}%
\bibitem [{\citenamefont {Cover}(1999)}]{Cover1999}%
  \BibitemOpen
  \bibfield  {author} {\bibinfo {author} {\bibfnamefont {T.~M.}\ \bibnamefont
  {Cover}},\ }\href@noop {} {\emph {\bibinfo {title} {{Elements of information
  theory}}}}\ (\bibinfo  {publisher} {John Wiley {\&} Sons},\ \bibinfo {year}
  {1999})\BibitemShut {NoStop}%
\bibitem [{\citenamefont {Shim}\ \emph {et~al.}(2019)\citenamefont {Shim},
  \citenamefont {Fan}, \citenamefont {Johnson},\ and\ \citenamefont
  {Miller}}]{Shim2019}%
  \BibitemOpen
  \bibfield  {author} {\bibinfo {author} {\bibfnamefont {H.}~\bibnamefont
  {Shim}}, \bibinfo {author} {\bibfnamefont {L.}~\bibnamefont {Fan}}, \bibinfo
  {author} {\bibfnamefont {S.~G.}\ \bibnamefont {Johnson}},\ and\ \bibinfo
  {author} {\bibfnamefont {O.~D.}\ \bibnamefont {Miller}},\ }\bibfield  {title}
  {\bibinfo {title} {{Fundamental Limits to Near-Field Optical Response over
  Any Bandwidth}},\ }\href {https://doi.org/10.1103/PhysRevX.9.011043}
  {\bibfield  {journal} {\bibinfo  {journal} {Phys. Rev. X}\ }\textbf {\bibinfo
  {volume} {9}},\ \bibinfo {pages} {11043} (\bibinfo {year}
  {2019})}\BibitemShut {NoStop}%
\bibitem [{\citenamefont {Callen}(1985)}]{Callen1985}%
  \BibitemOpen
  \bibfield  {author} {\bibinfo {author} {\bibfnamefont {H.~B.}\ \bibnamefont
  {Callen}},\ }\href@noop {} {\emph {\bibinfo {title} {{Thermodynamics and an
  Introduction to Thermostatistics}}}}\ (\bibinfo  {publisher} {John Wiley {\&}
  Sons},\ \bibinfo {year} {1985})\BibitemShut {NoStop}%
\bibitem [{\citenamefont {Shockley}\ and\ \citenamefont
  {Queisser}(1961)}]{Shockley1961}%
  \BibitemOpen
  \bibfield  {author} {\bibinfo {author} {\bibfnamefont {W.}~\bibnamefont
  {Shockley}}\ and\ \bibinfo {author} {\bibfnamefont {H.~J.}\ \bibnamefont
  {Queisser}},\ }\bibfield  {title} {\bibinfo {title} {{Detailed balance limit
  of efficiency of p-n junction solar cells}},\ }\href
  {https://doi.org/10.1063/1.1736034} {\bibfield  {journal} {\bibinfo
  {journal} {J. Appl. Phys.}\ }\textbf {\bibinfo {volume} {32}},\ \bibinfo
  {pages} {510} (\bibinfo {year} {1961})}\BibitemShut {NoStop}%
\bibitem [{\citenamefont {Bergman}(1981)}]{Bergman1981}%
  \BibitemOpen
  \bibfield  {author} {\bibinfo {author} {\bibfnamefont {D.~J.}\ \bibnamefont
  {Bergman}},\ }\bibfield  {title} {\bibinfo {title} {{Bounds for the complex
  dielectric constant of a two-component composite material}},\ }\href
  {https://doi.org/10.1103/PhysRevB.23.3058} {\bibfield  {journal} {\bibinfo
  {journal} {Phys. Rev. B}\ }\textbf {\bibinfo {volume} {23}},\ \bibinfo
  {pages} {3058} (\bibinfo {year} {1981})}\BibitemShut {NoStop}%
\bibitem [{\citenamefont {Milton}(1981)}]{Milton1981}%
  \BibitemOpen
  \bibfield  {author} {\bibinfo {author} {\bibfnamefont {G.~W.}\ \bibnamefont
  {Milton}},\ }\bibfield  {title} {\bibinfo {title} {{Bounds on the complex
  permittivity of a two-component composite material}},\ }\href
  {https://doi.org/10.1063/1.329385} {\bibfield  {journal} {\bibinfo  {journal}
  {J. Appl. Phys.}\ }\textbf {\bibinfo {volume} {52}},\ \bibinfo {pages} {5286}
  (\bibinfo {year} {1981})}\BibitemShut {NoStop}%
\bibitem [{\citenamefont {Milton}(2002)}]{Milton2002}%
  \BibitemOpen
  \bibfield  {author} {\bibinfo {author} {\bibfnamefont {G.~W.}\ \bibnamefont
  {Milton}},\ }\href@noop {} {\emph {\bibinfo {title} {{The Theory of
  Composites}}}}\ (\bibinfo  {publisher} {Cambridge University Press},\
  \bibinfo {year} {2002})\BibitemShut {NoStop}%
\bibitem [{\citenamefont {Wheeler}(1947)}]{Wheeler1947}%
  \BibitemOpen
  \bibfield  {author} {\bibinfo {author} {\bibfnamefont {H.~A.}\ \bibnamefont
  {Wheeler}},\ }\bibfield  {title} {\bibinfo {title} {{Fundamental Limitations
  of Small Antennas}},\ }\href {https://doi.org/10.1109/JRPROC.1947.226199}
  {\bibfield  {journal} {\bibinfo  {journal} {Proc. IRE}\ }\textbf {\bibinfo
  {volume} {35}},\ \bibinfo {pages} {1479} (\bibinfo {year}
  {1947})}\BibitemShut {NoStop}%
\bibitem [{\citenamefont {Chu}(1948)}]{Chu1948}%
  \BibitemOpen
  \bibfield  {author} {\bibinfo {author} {\bibfnamefont {L.~J.}\ \bibnamefont
  {Chu}},\ }\bibfield  {title} {\bibinfo {title} {{Physical Limitations of
  Omni-Directional Antennas}},\ }\href {https://doi.org/10.1063/1.1715038}
  {\bibfield  {journal} {\bibinfo  {journal} {J. Appl. Phys.}\ }\textbf
  {\bibinfo {volume} {19}},\ \bibinfo {pages} {1163} (\bibinfo {year}
  {1948})}\BibitemShut {NoStop}%
\bibitem [{\citenamefont {Gordon}(1963)}]{gordon_1963}%
  \BibitemOpen
  \bibfield  {author} {\bibinfo {author} {\bibfnamefont {R.~G.}\ \bibnamefont
  {Gordon}},\ }\bibfield  {title} {\bibinfo {title} {Three sum rules for total
  optical absorption cross sections},\ }\href
  {https://doi.org/10.1063/1.1776946} {\bibfield  {journal} {\bibinfo
  {journal} {The Journal of Chemical Physics}\ }\textbf {\bibinfo {volume}
  {38}},\ \bibinfo {pages} {1724} (\bibinfo {year} {1963})}\BibitemShut
  {NoStop}%
\bibitem [{\citenamefont {Purcell}(1969)}]{purcell_1969}%
  \BibitemOpen
  \bibfield  {author} {\bibinfo {author} {\bibfnamefont {E.~M.}\ \bibnamefont
  {Purcell}},\ }\bibfield  {title} {\bibinfo {title} {On the absorption and
  emission of light by interstellar grains},\ }\href
  {https://doi.org/10.1086/150207} {\bibfield  {journal} {\bibinfo  {journal}
  {The Astrophysical Journal}\ }\textbf {\bibinfo {volume} {158}},\ \bibinfo
  {pages} {433} (\bibinfo {year} {1969})}\BibitemShut {NoStop}%
\bibitem [{\citenamefont {Mckellar}\ \emph {et~al.}(1982)\citenamefont
  {Mckellar}, \citenamefont {Box},\ and\ \citenamefont
  {Bohren}}]{mckellar_box_bohren_1982}%
  \BibitemOpen
  \bibfield  {author} {\bibinfo {author} {\bibfnamefont {B.~H.~J.}\
  \bibnamefont {Mckellar}}, \bibinfo {author} {\bibfnamefont {M.~A.}\
  \bibnamefont {Box}},\ and\ \bibinfo {author} {\bibfnamefont {C.~F.}\
  \bibnamefont {Bohren}},\ }\bibfield  {title} {\bibinfo {title} {Sum rules for
  optical scattering amplitudes},\ }\href
  {https://doi.org/10.1364/josa.72.000535} {\bibfield  {journal} {\bibinfo
  {journal} {Journal of the Optical Society of America}\ }\textbf {\bibinfo
  {volume} {72}},\ \bibinfo {pages} {535} (\bibinfo {year} {1982})}\BibitemShut
  {NoStop}%
\bibitem [{\citenamefont {Sohl}\ \emph {et~al.}(2007)\citenamefont {Sohl},
  \citenamefont {Gustafsson},\ and\ \citenamefont
  {Kristensson}}]{sohl_gustafsson_kristensson_2007}%
  \BibitemOpen
  \bibfield  {author} {\bibinfo {author} {\bibfnamefont {C.}~\bibnamefont
  {Sohl}}, \bibinfo {author} {\bibfnamefont {M.}~\bibnamefont {Gustafsson}},\
  and\ \bibinfo {author} {\bibfnamefont {G.}~\bibnamefont {Kristensson}},\
  }\bibfield  {title} {\bibinfo {title} {Physical limitations on broadband
  scattering by heterogeneous obstacles},\ }\href
  {https://doi.org/10.1088/1751-8113/40/36/015} {\bibfield  {journal} {\bibinfo
   {journal} {Journal of Physics A: Mathematical and Theoretical}\ }\textbf
  {\bibinfo {volume} {40}},\ \bibinfo {pages} {11165} (\bibinfo {year}
  {2007})}\BibitemShut {NoStop}%
\bibitem [{\citenamefont {Miller}\ \emph {et~al.}(2015)\citenamefont {Miller},
  \citenamefont {Johnson},\ and\ \citenamefont
  {Rodriguez}}]{miller_johnson_rodriguez_2015}%
  \BibitemOpen
  \bibfield  {author} {\bibinfo {author} {\bibfnamefont {O.~D.}\ \bibnamefont
  {Miller}}, \bibinfo {author} {\bibfnamefont {S.~G.}\ \bibnamefont
  {Johnson}},\ and\ \bibinfo {author} {\bibfnamefont {A.~W.}\ \bibnamefont
  {Rodriguez}},\ }\bibfield  {title} {\bibinfo {title} {Shape-independent
  limits to near-field radiative heat transfer},\ }\href
  {https://doi.org/10.1103/physrevlett.115.204302} {\bibfield  {journal}
  {\bibinfo  {journal} {Physical Review Letters}\ }\textbf {\bibinfo {volume}
  {115}},\ \bibinfo {pages} {204302} (\bibinfo {year} {2015})}\BibitemShut
  {NoStop}%
\bibitem [{\citenamefont {Hugonin}\ \emph {et~al.}(2015)\citenamefont
  {Hugonin}, \citenamefont {Besbes},\ and\ \citenamefont
  {Ben-Abdallah}}]{hugonin_besbes_ben-abdallah_2015}%
  \BibitemOpen
  \bibfield  {author} {\bibinfo {author} {\bibfnamefont {J.-P.}\ \bibnamefont
  {Hugonin}}, \bibinfo {author} {\bibfnamefont {M.}~\bibnamefont {Besbes}},\
  and\ \bibinfo {author} {\bibfnamefont {P.}~\bibnamefont {Ben-Abdallah}},\
  }\bibfield  {title} {\bibinfo {title} {Fundamental limits for light
  absorption and scattering induced by cooperative electromagnetic
  interactions},\ }\href {https://doi.org/10.1103/physrevb.91.180202}
  {\bibfield  {journal} {\bibinfo  {journal} {Physical Review B}\ }\textbf
  {\bibinfo {volume} {91}},\ \bibinfo {pages} {180202(R)} (\bibinfo {year}
  {2015})}\BibitemShut {NoStop}%
\bibitem [{\citenamefont {Miller}\ \emph {et~al.}(2016)\citenamefont {Miller},
  \citenamefont {Polimeridis}, \citenamefont {Reid}, \citenamefont {Hsu},
  \citenamefont {Delacy}, \citenamefont {Joannopoulos}, \citenamefont {Solja{\v
  c}i\`c},\ and\ \citenamefont
  {Johnson}}]{miller_polimeridis_reid_hsu_delacy_joannopoulos_soljacic_johnson_2016}%
  \BibitemOpen
  \bibfield  {author} {\bibinfo {author} {\bibfnamefont {O.~D.}\ \bibnamefont
  {Miller}}, \bibinfo {author} {\bibfnamefont {A.~G.}\ \bibnamefont
  {Polimeridis}}, \bibinfo {author} {\bibfnamefont {M.~T.~H.}\ \bibnamefont
  {Reid}}, \bibinfo {author} {\bibfnamefont {C.~W.}\ \bibnamefont {Hsu}},
  \bibinfo {author} {\bibfnamefont {B.~G.}\ \bibnamefont {Delacy}}, \bibinfo
  {author} {\bibfnamefont {J.~D.}\ \bibnamefont {Joannopoulos}}, \bibinfo
  {author} {\bibfnamefont {M.}~\bibnamefont {Solja{\v c}i\`c}},\ and\ \bibinfo
  {author} {\bibfnamefont {S.~G.}\ \bibnamefont {Johnson}},\ }\bibfield
  {title} {\bibinfo {title} {Fundamental limits to optical response in
  absorptive systems},\ }\href {https://doi.org/10.1364/oe.24.003329}
  {\bibfield  {journal} {\bibinfo  {journal} {Optics Express}\ }\textbf
  {\bibinfo {volume} {24}},\ \bibinfo {pages} {3329} (\bibinfo {year}
  {2016})}\BibitemShut {NoStop}%
\bibitem [{\citenamefont {Miller}\ \emph {et~al.}(2017)\citenamefont {Miller},
  \citenamefont {Ilic}, \citenamefont {Christensen}, \citenamefont {Reid},
  \citenamefont {Atwater}, \citenamefont {Joannopoulos}, \citenamefont
  {Solja{\v c}i\`c},\ and\ \citenamefont
  {Johnson}}]{miller_ilic_christensen_reid_atwater_joannopoulos_soljacic_johnson_2017}%
  \BibitemOpen
  \bibfield  {author} {\bibinfo {author} {\bibfnamefont {O.~D.}\ \bibnamefont
  {Miller}}, \bibinfo {author} {\bibfnamefont {O.}~\bibnamefont {Ilic}},
  \bibinfo {author} {\bibfnamefont {T.}~\bibnamefont {Christensen}}, \bibinfo
  {author} {\bibfnamefont {M.~T.~H.}\ \bibnamefont {Reid}}, \bibinfo {author}
  {\bibfnamefont {H.~A.}\ \bibnamefont {Atwater}}, \bibinfo {author}
  {\bibfnamefont {J.~D.}\ \bibnamefont {Joannopoulos}}, \bibinfo {author}
  {\bibfnamefont {M.}~\bibnamefont {Solja{\v c}i\`c}},\ and\ \bibinfo {author}
  {\bibfnamefont {S.~G.}\ \bibnamefont {Johnson}},\ }\bibfield  {title}
  {\bibinfo {title} {Limits to the optical response of graphene and
  two-dimensional materials},\ }\href
  {https://doi.org/10.1021/acs.nanolett.7b02007} {\bibfield  {journal}
  {\bibinfo  {journal} {Nano Letters}\ }\textbf {\bibinfo {volume} {17}},\
  \bibinfo {pages} {5408} (\bibinfo {year} {2017})}\BibitemShut {NoStop}%
\bibitem [{\citenamefont {Sanders}\ and\ \citenamefont
  {Manjavacas}(2018)}]{sanders_manjavacas_2018}%
  \BibitemOpen
  \bibfield  {author} {\bibinfo {author} {\bibfnamefont {S.}~\bibnamefont
  {Sanders}}\ and\ \bibinfo {author} {\bibfnamefont {A.}~\bibnamefont
  {Manjavacas}},\ }\bibfield  {title} {\bibinfo {title} {Analysis of the limits
  of the local density of photonic states near nanostructures},\ }\bibfield
  {journal} {\bibinfo  {journal} {ACS Photonics}\ }\href
  {https://doi.org/10.1021/acsphotonics.8b00225} {10.1021/acsphotonics.8b00225}
  (\bibinfo {year} {2018})\BibitemShut {NoStop}%
\bibitem [{\citenamefont {Yang}\ \emph {et~al.}(2017)\citenamefont {Yang},
  \citenamefont {Miller}, \citenamefont {Christensen}, \citenamefont
  {Joannopoulos},\ and\ \citenamefont {Solja{\v
  c}i\`c}}]{yang_miller_christensen_joannopoulos_soljacic_2017}%
  \BibitemOpen
  \bibfield  {author} {\bibinfo {author} {\bibfnamefont {Y.}~\bibnamefont
  {Yang}}, \bibinfo {author} {\bibfnamefont {O.~D.}\ \bibnamefont {Miller}},
  \bibinfo {author} {\bibfnamefont {T.}~\bibnamefont {Christensen}}, \bibinfo
  {author} {\bibfnamefont {J.~D.}\ \bibnamefont {Joannopoulos}},\ and\ \bibinfo
  {author} {\bibfnamefont {M.}~\bibnamefont {Solja{\v c}i\`c}},\ }\bibfield
  {title} {\bibinfo {title} {Low-loss plasmonic dielectric nanoresonators},\
  }\href {https://doi.org/10.1021/acs.nanolett.7b00852} {\bibfield  {journal}
  {\bibinfo  {journal} {Nano Letters}\ }\textbf {\bibinfo {volume} {17}},\
  \bibinfo {pages} {3238} (\bibinfo {year} {2017})}\BibitemShut {NoStop}%
\bibitem [{\citenamefont {Yang}\ \emph {et~al.}(2018)\citenamefont {Yang},
  \citenamefont {Massuda}, \citenamefont {Roques-Carmes}, \citenamefont {Kooi},
  \citenamefont {Christensen}, \citenamefont {Johnson}, \citenamefont
  {Joannopoulos}, \citenamefont {Miller}, \citenamefont {Kaminer},\ and\
  \citenamefont {Solja{\v{c}}i{\'{c}}}}]{Yang2018}%
  \BibitemOpen
  \bibfield  {author} {\bibinfo {author} {\bibfnamefont {Y.}~\bibnamefont
  {Yang}}, \bibinfo {author} {\bibfnamefont {A.}~\bibnamefont {Massuda}},
  \bibinfo {author} {\bibfnamefont {C.}~\bibnamefont {Roques-Carmes}}, \bibinfo
  {author} {\bibfnamefont {S.~E.}\ \bibnamefont {Kooi}}, \bibinfo {author}
  {\bibfnamefont {T.}~\bibnamefont {Christensen}}, \bibinfo {author}
  {\bibfnamefont {S.~G.}\ \bibnamefont {Johnson}}, \bibinfo {author}
  {\bibfnamefont {J.~D.}\ \bibnamefont {Joannopoulos}}, \bibinfo {author}
  {\bibfnamefont {O.~D.}\ \bibnamefont {Miller}}, \bibinfo {author}
  {\bibfnamefont {I.}~\bibnamefont {Kaminer}},\ and\ \bibinfo {author}
  {\bibfnamefont {M.}~\bibnamefont {Solja{\v{c}}i{\'{c}}}},\ }\bibfield
  {title} {\bibinfo {title} {Maximal spontaneous photon emission and energy
  loss from free electrons},\ }\href
  {https://doi.org/10.1038/s41567-018-0180-2} {\bibfield  {journal} {\bibinfo
  {journal} {Nature Physics}\ }\textbf {\bibinfo {volume} {14}},\ \bibinfo
  {pages} {894} (\bibinfo {year} {2018})}\BibitemShut {NoStop}%
\bibitem [{\citenamefont {Michon}\ \emph {et~al.}(2019)\citenamefont {Michon},
  \citenamefont {Benzaouia}, \citenamefont {Yao}, \citenamefont {Miller},\ and\
  \citenamefont {Johnson}}]{Michon2019}%
  \BibitemOpen
  \bibfield  {author} {\bibinfo {author} {\bibfnamefont {J.}~\bibnamefont
  {Michon}}, \bibinfo {author} {\bibfnamefont {M.}~\bibnamefont {Benzaouia}},
  \bibinfo {author} {\bibfnamefont {W.}~\bibnamefont {Yao}}, \bibinfo {author}
  {\bibfnamefont {O.~D.}\ \bibnamefont {Miller}},\ and\ \bibinfo {author}
  {\bibfnamefont {S.~G.}\ \bibnamefont {Johnson}},\ }\bibfield  {title}
  {\bibinfo {title} {Limits to surface-enhanced raman scattering near
  arbitrary-shape scatterers},\ }\href {https://doi.org/10.1364/OE.27.035189}
  {\bibfield  {journal} {\bibinfo  {journal} {Opt. Express}\ }\textbf {\bibinfo
  {volume} {27}},\ \bibinfo {pages} {35189} (\bibinfo {year} {2019})},\ \Eprint
  {https://arxiv.org/abs/1909.00202} {1909.00202} \BibitemShut {NoStop}%
\bibitem [{\citenamefont {Zhang}\ \emph {et~al.}(2019)\citenamefont {Zhang},
  \citenamefont {Hsu},\ and\ \citenamefont {Miller}}]{Zhang2019}%
  \BibitemOpen
  \bibfield  {author} {\bibinfo {author} {\bibfnamefont {H.}~\bibnamefont
  {Zhang}}, \bibinfo {author} {\bibfnamefont {C.~W.}\ \bibnamefont {Hsu}},\
  and\ \bibinfo {author} {\bibfnamefont {O.~D.}\ \bibnamefont {Miller}},\
  }\bibfield  {title} {\bibinfo {title} {Scattering concentration bounds:
  Brightness theorems for waves},\ }\href
  {https://doi.org/10.1364/OPTICA.6.001321} {\bibfield  {journal} {\bibinfo
  {journal} {Optica}\ }\textbf {\bibinfo {volume} {6}},\ \bibinfo {pages}
  {1321} (\bibinfo {year} {2019})},\ \Eprint {https://arxiv.org/abs/1810.02727}
  {1810.02727} \BibitemShut {NoStop}%
\bibitem [{\citenamefont {Molesky}\ \emph {et~al.}(2019)\citenamefont
  {Molesky}, \citenamefont {Jin}, \citenamefont {Venkataram},\ and\
  \citenamefont {Rodriguez}}]{Molesky2019a}%
  \BibitemOpen
  \bibfield  {author} {\bibinfo {author} {\bibfnamefont {S.}~\bibnamefont
  {Molesky}}, \bibinfo {author} {\bibfnamefont {W.}~\bibnamefont {Jin}},
  \bibinfo {author} {\bibfnamefont {P.~S.}\ \bibnamefont {Venkataram}},\ and\
  \bibinfo {author} {\bibfnamefont {A.~W.}\ \bibnamefont {Rodriguez}},\
  }\bibfield  {title} {\bibinfo {title} {{T Operator Bounds on Angle-Integrated
  Absorption and Thermal Radiation for Arbitrary Objects}},\ }\href
  {https://doi.org/10.1103/PhysRevLett.123.257401} {\bibfield  {journal}
  {\bibinfo  {journal} {Phys. Rev. Lett.}\ }\textbf {\bibinfo {volume} {123}},\
  \bibinfo {pages} {257401} (\bibinfo {year} {2019})}\BibitemShut {NoStop}%
\bibitem [{\citenamefont {Shim}\ \emph
  {et~al.}(2020{\natexlab{a}})\citenamefont {Shim}, \citenamefont {Chung},\
  and\ \citenamefont {Miller}}]{ShimChung2020}%
  \BibitemOpen
  \bibfield  {author} {\bibinfo {author} {\bibfnamefont {H.}~\bibnamefont
  {Shim}}, \bibinfo {author} {\bibfnamefont {H.}~\bibnamefont {Chung}},\ and\
  \bibinfo {author} {\bibfnamefont {O.~D.}\ \bibnamefont {Miller}},\ }\bibfield
   {title} {\bibinfo {title} {Maximal free-space concentration of
  electromagnetic waves},\ }\href
  {https://doi.org/10.1103/PhysRevApplied.14.014007} {\bibfield  {journal}
  {\bibinfo  {journal} {Physical Review Applied}\ }\textbf {\bibinfo {volume}
  {14}},\ \bibinfo {pages} {014007} (\bibinfo {year} {2020}{\natexlab{a}})},\
  \Eprint {https://arxiv.org/abs/1905.10500} {1905.10500} \BibitemShut
  {NoStop}%
\bibitem [{\citenamefont {Shim}\ \emph
  {et~al.}(2020{\natexlab{b}})\citenamefont {Shim}, \citenamefont {Kuang},\
  and\ \citenamefont {Miller}}]{Shim2020}%
  \BibitemOpen
  \bibfield  {author} {\bibinfo {author} {\bibfnamefont {H.}~\bibnamefont
  {Shim}}, \bibinfo {author} {\bibfnamefont {Z.}~\bibnamefont {Kuang}},\ and\
  \bibinfo {author} {\bibfnamefont {O.~D.}\ \bibnamefont {Miller}},\ }\bibfield
   {title} {\bibinfo {title} {Optical materials for maximal nanophotonic
  response ({I}nvited)},\ }\href {https://doi.org/10.1364/OME.396419}
  {\bibfield  {journal} {\bibinfo  {journal} {Optical Materials Express}\
  }\textbf {\bibinfo {volume} {10}},\ \bibinfo {pages} {1561} (\bibinfo {year}
  {2020}{\natexlab{b}})},\ \Eprint {https://arxiv.org/abs/2004.13132}
  {arXiv:2004.13132} \BibitemShut {NoStop}%
\bibitem [{\citenamefont {Molesky}\ \emph
  {et~al.}(2020{\natexlab{a}})\citenamefont {Molesky}, \citenamefont {Chao},
  \citenamefont {Jin},\ and\ \citenamefont {Rodriguez}}]{Molesky2020a}%
  \BibitemOpen
  \bibfield  {author} {\bibinfo {author} {\bibfnamefont {S.}~\bibnamefont
  {Molesky}}, \bibinfo {author} {\bibfnamefont {P.}~\bibnamefont {Chao}},
  \bibinfo {author} {\bibfnamefont {W.}~\bibnamefont {Jin}},\ and\ \bibinfo
  {author} {\bibfnamefont {A.~W.}\ \bibnamefont {Rodriguez}},\ }\bibfield
  {title} {\bibinfo {title} {{Global T operator bounds on electromagnetic
  scattering: Upper bounds on far-field cross sections}},\ }\href
  {https://doi.org/10.1103/physrevresearch.2.033172} {\bibfield  {journal}
  {\bibinfo  {journal} {Phys. Rev. Res.}\ }\textbf {\bibinfo {volume} {2}},\
  \bibinfo {pages} {033172} (\bibinfo {year} {2020}{\natexlab{a}})}\BibitemShut
  {NoStop}%
\bibitem [{\citenamefont {Gustafsson}\ \emph {et~al.}(2019)\citenamefont
  {Gustafsson}, \citenamefont {Schab}, \citenamefont {Jelinek},\ and\
  \citenamefont {Capek}}]{Gustafsson2019}%
  \BibitemOpen
  \bibfield  {author} {\bibinfo {author} {\bibfnamefont {M.}~\bibnamefont
  {Gustafsson}}, \bibinfo {author} {\bibfnamefont {K.}~\bibnamefont {Schab}},
  \bibinfo {author} {\bibfnamefont {L.}~\bibnamefont {Jelinek}},\ and\ \bibinfo
  {author} {\bibfnamefont {M.}~\bibnamefont {Capek}},\ }\bibfield  {title}
  {\bibinfo {title} {{Upper bounds on absorption and scattering}},\ }\href
  {http://arxiv.org/abs/1912.06699} {\bibfield  {journal} {\bibinfo  {journal}
  {arXiv Prepr. arXiv1912.06699}\ } (\bibinfo {year} {2019})},\ \Eprint
  {https://arxiv.org/abs/1912.06699} {arXiv:1912.06699} \BibitemShut {NoStop}%
\bibitem [{\citenamefont {Molesky}\ \emph
  {et~al.}(2020{\natexlab{b}})\citenamefont {Molesky}, \citenamefont
  {Venkataram}, \citenamefont {Jin},\ and\ \citenamefont
  {Rodriguez}}]{Molesky2020b}%
  \BibitemOpen
  \bibfield  {author} {\bibinfo {author} {\bibfnamefont {S.}~\bibnamefont
  {Molesky}}, \bibinfo {author} {\bibfnamefont {P.~S.}\ \bibnamefont
  {Venkataram}}, \bibinfo {author} {\bibfnamefont {W.}~\bibnamefont {Jin}},\
  and\ \bibinfo {author} {\bibfnamefont {A.~W.}\ \bibnamefont {Rodriguez}},\
  }\bibfield  {title} {\bibinfo {title} {{Fundamental limits to radiative heat
  transfer: Theory}},\ }\href {https://doi.org/10.1103/PhysRevB.101.035408}
  {\bibfield  {journal} {\bibinfo  {journal} {Phys. Rev. B}\ }\textbf {\bibinfo
  {volume} {101}},\ \bibinfo {pages} {35408} (\bibinfo {year}
  {2020}{\natexlab{b}})},\ \Eprint {https://arxiv.org/abs/1907.03000}
  {arXiv:1907.03000} \BibitemShut {NoStop}%
\bibitem [{\citenamefont {Kuang}\ \emph {et~al.}(2020)\citenamefont {Kuang},
  \citenamefont {Zhang},\ and\ \citenamefont {Miller}}]{Kuang2020subm}%
  \BibitemOpen
  \bibfield  {author} {\bibinfo {author} {\bibfnamefont {Z.}~\bibnamefont
  {Kuang}}, \bibinfo {author} {\bibfnamefont {L.}~\bibnamefont {Zhang}},\ and\
  \bibinfo {author} {\bibfnamefont {O.~D.}\ \bibnamefont {Miller}},\ }\bibfield
   {title} {\bibinfo {title} {Maximal single-frequency electromagnetic
  response},\ }\href@noop {} {\bibfield  {journal} {\bibinfo  {journal}
  {arXiv:2002.00521}\ } (\bibinfo {year} {2020})},\ \Eprint
  {https://arxiv.org/abs/2002.00521} {arXiv:2002.00521} \BibitemShut {NoStop}%
\bibitem [{\citenamefont {Angeris}\ \emph {et~al.}(2019)\citenamefont
  {Angeris}, \citenamefont {Vuckovic},\ and\ \citenamefont
  {Boyd}}]{Angeris2019}%
  \BibitemOpen
  \bibfield  {author} {\bibinfo {author} {\bibfnamefont {G.}~\bibnamefont
  {Angeris}}, \bibinfo {author} {\bibfnamefont {J.}~\bibnamefont {Vuckovic}},\
  and\ \bibinfo {author} {\bibfnamefont {S.~P.}\ \bibnamefont {Boyd}},\
  }\bibfield  {title} {\bibinfo {title} {{Computational Bounds for Photonic
  Design}},\ }\href {https://doi.org/10.1021/acsphotonics.9b00154} {\bibfield
  {journal} {\bibinfo  {journal} {ACS Photonics}\ }\textbf {\bibinfo {volume}
  {6}},\ \bibinfo {pages} {1232} (\bibinfo {year} {2019})}\BibitemShut
  {NoStop}%
\bibitem [{\citenamefont {Trivedi}\ \emph {et~al.}(2020)\citenamefont
  {Trivedi}, \citenamefont {Angeris}, \citenamefont {Su}, \citenamefont {Boyd},
  \citenamefont {Fan},\ and\ \citenamefont {Vuckovic}}]{Trivedi2020}%
  \BibitemOpen
  \bibfield  {author} {\bibinfo {author} {\bibfnamefont {R.}~\bibnamefont
  {Trivedi}}, \bibinfo {author} {\bibfnamefont {G.}~\bibnamefont {Angeris}},
  \bibinfo {author} {\bibfnamefont {L.}~\bibnamefont {Su}}, \bibinfo {author}
  {\bibfnamefont {S.}~\bibnamefont {Boyd}}, \bibinfo {author} {\bibfnamefont
  {S.}~\bibnamefont {Fan}},\ and\ \bibinfo {author} {\bibfnamefont
  {J.}~\bibnamefont {Vuckovic}},\ }\bibfield  {title} {\bibinfo {title}
  {{Fundamental bounds for scattering from absorptionless electromagnetic
  structures}},\ }\href {http://arxiv.org/abs/2003.00374} {\bibfield  {journal}
  {\bibinfo  {journal} {arXiv Prepr. arXiv2003.00374}\ ,\ \bibinfo {pages} {1}}
  (\bibinfo {year} {2020})},\ \Eprint {https://arxiv.org/abs/2003.00374}
  {arXiv:2003.00374} \BibitemShut {NoStop}%
\bibitem [{\citenamefont {Yu}\ \emph {et~al.}(2010)\citenamefont {Yu},
  \citenamefont {Raman},\ and\ \citenamefont {Fan}}]{Yu2010}%
  \BibitemOpen
  \bibfield  {author} {\bibinfo {author} {\bibfnamefont {Z.}~\bibnamefont
  {Yu}}, \bibinfo {author} {\bibfnamefont {A.}~\bibnamefont {Raman}},\ and\
  \bibinfo {author} {\bibfnamefont {S.}~\bibnamefont {Fan}},\ }\bibfield
  {title} {\bibinfo {title} {{Fundamental limit of nanophotonic light trapping
  in solar cells}},\ }\href {https://doi.org/10.1073/pnas.1008296107}
  {\bibfield  {journal} {\bibinfo  {journal} {Proc. Natl. Acad. Sci. U. S. A.}\
  }\textbf {\bibinfo {volume} {107}},\ \bibinfo {pages} {17491} (\bibinfo
  {year} {2010})}\BibitemShut {NoStop}%
\bibitem [{\citenamefont {Presutti}\ and\ \citenamefont
  {Monticone}(2020)}]{Presutti2020}%
  \BibitemOpen
  \bibfield  {author} {\bibinfo {author} {\bibfnamefont {F.}~\bibnamefont
  {Presutti}}\ and\ \bibinfo {author} {\bibfnamefont {F.}~\bibnamefont
  {Monticone}},\ }\bibfield  {title} {\bibinfo {title} {{Focusing on bandwidth:
  achromatic metalens limits}},\ }\href {https://doi.org/10.1364/optica.389404}
  {\bibfield  {journal} {\bibinfo  {journal} {Optica}\ }\textbf {\bibinfo
  {volume} {7}},\ \bibinfo {pages} {624} (\bibinfo {year} {2020})},\ \Eprint
  {https://arxiv.org/abs/2001.10899} {arXiv:2001.10899} \BibitemShut {NoStop}%
\bibitem [{\citenamefont {Chung}\ and\ \citenamefont
  {Miller}(2020)}]{Chung2020}%
  \BibitemOpen
  \bibfield  {author} {\bibinfo {author} {\bibfnamefont {H.}~\bibnamefont
  {Chung}}\ and\ \bibinfo {author} {\bibfnamefont {O.~D.}\ \bibnamefont
  {Miller}},\ }\bibfield  {title} {\bibinfo {title} {High-{NA} achromatic
  metalenses by inverse design},\ }\href {https://doi.org/10.1364/OE.385440}
  {\bibfield  {journal} {\bibinfo  {journal} {Optics Express}\ }\textbf
  {\bibinfo {volume} {28}},\ \bibinfo {pages} {6945} (\bibinfo {year}
  {2020})},\ \Eprint {https://arxiv.org/abs/1905.09213} {1905.09213}
  \BibitemShut {NoStop}%
\bibitem [{\citenamefont {Laurent}\ and\ \citenamefont
  {Rendl}(2005)}]{Laurent2005}%
  \BibitemOpen
  \bibfield  {author} {\bibinfo {author} {\bibfnamefont {M.}~\bibnamefont
  {Laurent}}\ and\ \bibinfo {author} {\bibfnamefont {F.}~\bibnamefont
  {Rendl}},\ }\bibfield  {title} {\bibinfo {title} {{Semidefinite Programming
  and Integer Programming}},\ }\href
  {https://doi.org/10.1016/S0927-0507(05)12008-8} {\bibfield  {journal}
  {\bibinfo  {journal} {Handbooks Oper. Res. Manag. Sci.}\ }\textbf {\bibinfo
  {volume} {12}},\ \bibinfo {pages} {393} (\bibinfo {year} {2005})}\BibitemShut
  {NoStop}%
\bibitem [{\citenamefont {Luo}\ \emph {et~al.}(2010)\citenamefont {Luo},
  \citenamefont {Ma}, \citenamefont {So}, \citenamefont {Ye},\ and\
  \citenamefont {Zhang}}]{Luo2010}%
  \BibitemOpen
  \bibfield  {author} {\bibinfo {author} {\bibfnamefont {Z.~Q.}\ \bibnamefont
  {Luo}}, \bibinfo {author} {\bibfnamefont {W.~K.}\ \bibnamefont {Ma}},
  \bibinfo {author} {\bibfnamefont {A.}~\bibnamefont {So}}, \bibinfo {author}
  {\bibfnamefont {Y.}~\bibnamefont {Ye}},\ and\ \bibinfo {author}
  {\bibfnamefont {S.}~\bibnamefont {Zhang}},\ }\bibfield  {title} {\bibinfo
  {title} {{Semidefinite relaxation of quadratic optimization problems}},\
  }\href {https://doi.org/10.1109/MSP.2010.936019} {\bibfield  {journal}
  {\bibinfo  {journal} {IEEE Signal Process. Mag.}\ }\textbf {\bibinfo {volume}
  {27}},\ \bibinfo {pages} {20} (\bibinfo {year} {2010})}\BibitemShut {NoStop}%
\bibitem [{\citenamefont {Jackson}(1999)}]{Jackson1999}%
  \BibitemOpen
  \bibfield  {author} {\bibinfo {author} {\bibfnamefont {J.~D.}\ \bibnamefont
  {Jackson}},\ }\href@noop {} {\emph {\bibinfo {title} {{Classical
  Electrodynamics, 3rd Ed.}}}}\ (\bibinfo  {publisher} {John Wiley {\&} Sons},\
  \bibinfo {year} {1999})\BibitemShut {NoStop}%
\bibitem [{\citenamefont {Newton}(1976)}]{Newton1976}%
  \BibitemOpen
  \bibfield  {author} {\bibinfo {author} {\bibfnamefont {R.~G.}\ \bibnamefont
  {Newton}},\ }\bibfield  {title} {\bibinfo {title} {{Optical theorem and
  beyond}},\ }\href {https://doi.org/10.1119/1.10324} {\bibfield  {journal}
  {\bibinfo  {journal} {Am. J. Phys.}\ }\textbf {\bibinfo {volume} {44}},\
  \bibinfo {pages} {639} (\bibinfo {year} {1976})}\BibitemShut {NoStop}%
\bibitem [{\citenamefont {Lytle}\ \emph {et~al.}(2005)\citenamefont {Lytle},
  \citenamefont {Carney}, \citenamefont {Schotland},\ and\ \citenamefont
  {Wolf}}]{Lytle2005}%
  \BibitemOpen
  \bibfield  {author} {\bibinfo {author} {\bibfnamefont {D.~R.}\ \bibnamefont
  {Lytle}}, \bibinfo {author} {\bibfnamefont {P.~S.}\ \bibnamefont {Carney}},
  \bibinfo {author} {\bibfnamefont {J.~C.}\ \bibnamefont {Schotland}},\ and\
  \bibinfo {author} {\bibfnamefont {E.}~\bibnamefont {Wolf}},\ }\bibfield
  {title} {\bibinfo {title} {{Generalized optical theorem for reflection,
  transmission, and extinction of power for electromagnetic fields}},\ }\href
  {https://doi.org/10.1103/PhysRevE.71.056610} {\bibfield  {journal} {\bibinfo
  {journal} {Phys. Rev. E}\ }\textbf {\bibinfo {volume} {71}},\ \bibinfo
  {pages} {056610} (\bibinfo {year} {2005})}\BibitemShut {NoStop}%
\bibitem [{\citenamefont {Jin}(2011)}]{Jin2011}%
  \BibitemOpen
  \bibfield  {author} {\bibinfo {author} {\bibfnamefont {J.-M.}\ \bibnamefont
  {Jin}},\ }\href@noop {} {\emph {\bibinfo {title} {{Theory and computation of
  electromagnetic fields}}}}\ (\bibinfo  {publisher} {John Wiley {\&} Sons},\
  \bibinfo {year} {2011})\BibitemShut {NoStop}%
\bibitem [{\citenamefont {Boyd}\ and\ \citenamefont
  {Vandenberghe}(2004)}]{Boyd2004}%
  \BibitemOpen
  \bibfield  {author} {\bibinfo {author} {\bibfnamefont {S.}~\bibnamefont
  {Boyd}}\ and\ \bibinfo {author} {\bibfnamefont {L.}~\bibnamefont
  {Vandenberghe}},\ }\href@noop {} {\emph {\bibinfo {title} {{Convex
  Optimization}}}}\ (\bibinfo  {publisher} {Cambridge University Press},\
  \bibinfo {address} {Cambridge, UK},\ \bibinfo {year} {2004})\BibitemShut
  {NoStop}%
\bibitem [{\citenamefont {Vandenberghe}\ and\ \citenamefont
  {Boyd}(1996)}]{Vandenberghe1996}%
  \BibitemOpen
  \bibfield  {author} {\bibinfo {author} {\bibfnamefont {L.}~\bibnamefont
  {Vandenberghe}}\ and\ \bibinfo {author} {\bibfnamefont {S.}~\bibnamefont
  {Boyd}},\ }\bibfield  {title} {\bibinfo {title} {{Semidefinite
  Programming}},\ }\href {https://doi.org/10.1137/1038003} {\bibfield
  {journal} {\bibinfo  {journal} {SIAM Rev.}\ }\textbf {\bibinfo {volume}
  {38}},\ \bibinfo {pages} {49} (\bibinfo {year} {1996})}\BibitemShut {NoStop}%
\bibitem [{\citenamefont {Goemans}\ and\ \citenamefont
  {Williamson}(1995)}]{Goemans1995}%
  \BibitemOpen
  \bibfield  {author} {\bibinfo {author} {\bibfnamefont {M.~X.}\ \bibnamefont
  {Goemans}}\ and\ \bibinfo {author} {\bibfnamefont {D.~P.}\ \bibnamefont
  {Williamson}},\ }\bibfield  {title} {\bibinfo {title} {{Improved
  approximation algorithms for maximum cut and satisfiability problems using
  semidefinite programming}},\ }\href {https://doi.org/10.1145/227683.227684}
  {\bibfield  {journal} {\bibinfo  {journal} {J. ACM}\ }\textbf {\bibinfo
  {volume} {42}},\ \bibinfo {pages} {1115} (\bibinfo {year}
  {1995})}\BibitemShut {NoStop}%
\bibitem [{\citenamefont {{Peng Hui Tan}}\ and\ \citenamefont
  {Rasmussen}(2001)}]{Tan2001}%
  \BibitemOpen
  \bibfield  {author} {\bibinfo {author} {\bibnamefont {{Peng Hui Tan}}}\ and\
  \bibinfo {author} {\bibfnamefont {L.}~\bibnamefont {Rasmussen}},\ }\bibfield
  {title} {\bibinfo {title} {{The application of semidefinite programming for
  detection in CDMA}},\ }\href {https://doi.org/10.1109/49.942507} {\bibfield
  {journal} {\bibinfo  {journal} {IEEE J. Sel. Areas Commun.}\ }\textbf
  {\bibinfo {volume} {19}},\ \bibinfo {pages} {1442} (\bibinfo {year}
  {2001})}\BibitemShut {NoStop}%
\bibitem [{\citenamefont {Biswas}\ \emph {et~al.}(2006)\citenamefont {Biswas},
  \citenamefont {Lian}, \citenamefont {Wang},\ and\ \citenamefont
  {Ye}}]{Biswas2006}%
  \BibitemOpen
  \bibfield  {author} {\bibinfo {author} {\bibfnamefont {P.}~\bibnamefont
  {Biswas}}, \bibinfo {author} {\bibfnamefont {T.~C.}\ \bibnamefont {Lian}},
  \bibinfo {author} {\bibfnamefont {T.~C.}\ \bibnamefont {Wang}},\ and\
  \bibinfo {author} {\bibfnamefont {Y.}~\bibnamefont {Ye}},\ }\bibfield
  {title} {\bibinfo {title} {{Semidefinite programming based algorithms for
  sensor network localization}},\ }\href
  {https://doi.org/10.1145/1149283.1149286} {\bibfield  {journal} {\bibinfo
  {journal} {ACM Trans. Sens. Networks}\ }\textbf {\bibinfo {volume} {2}},\
  \bibinfo {pages} {188} (\bibinfo {year} {2006})}\BibitemShut {NoStop}%
\bibitem [{\citenamefont {Gershman}\ \emph {et~al.}(2010)\citenamefont
  {Gershman}, \citenamefont {Sidiropoulos}, \citenamefont {Shahbazpanahi},
  \citenamefont {Bengtsson},\ and\ \citenamefont {Ottersten}}]{Gershman2010}%
  \BibitemOpen
  \bibfield  {author} {\bibinfo {author} {\bibfnamefont {A.~B.}\ \bibnamefont
  {Gershman}}, \bibinfo {author} {\bibfnamefont {N.~D.}\ \bibnamefont
  {Sidiropoulos}}, \bibinfo {author} {\bibfnamefont {S.}~\bibnamefont
  {Shahbazpanahi}}, \bibinfo {author} {\bibfnamefont {M.}~\bibnamefont
  {Bengtsson}},\ and\ \bibinfo {author} {\bibfnamefont {B.}~\bibnamefont
  {Ottersten}},\ }\bibfield  {title} {\bibinfo {title} {{Convex
  optimization-based beamforming}},\ }\href@noop {} {\bibfield  {journal}
  {\bibinfo  {journal} {IEEE Signal Process. Mag.}\ }\textbf {\bibinfo {volume}
  {27}},\ \bibinfo {pages} {62} (\bibinfo {year} {2010})}\BibitemShut {NoStop}%
\bibitem [{\citenamefont {Silva}\ \emph {et~al.}(2014)\citenamefont {Silva},
  \citenamefont {Monticone}, \citenamefont {Castaldi}, \citenamefont {Galdi},
  \citenamefont {Alu},\ and\ \citenamefont {Engheta}}]{Silva2014}%
  \BibitemOpen
  \bibfield  {author} {\bibinfo {author} {\bibfnamefont {A.}~\bibnamefont
  {Silva}}, \bibinfo {author} {\bibfnamefont {F.}~\bibnamefont {Monticone}},
  \bibinfo {author} {\bibfnamefont {G.}~\bibnamefont {Castaldi}}, \bibinfo
  {author} {\bibfnamefont {V.}~\bibnamefont {Galdi}}, \bibinfo {author}
  {\bibfnamefont {A.}~\bibnamefont {Alu}},\ and\ \bibinfo {author}
  {\bibfnamefont {N.}~\bibnamefont {Engheta}},\ }\bibfield  {title} {\bibinfo
  {title} {{Performing Mathematical Operations with Metamaterials}},\ }\href
  {https://doi.org/10.1126/science.1242818} {\bibfield  {journal} {\bibinfo
  {journal} {Science}\ }\textbf {\bibinfo {volume} {343}},\ \bibinfo {pages}
  {160} (\bibinfo {year} {2014})}\BibitemShut {NoStop}%
\bibitem [{\citenamefont {Pors}\ \emph {et~al.}(2015)\citenamefont {Pors},
  \citenamefont {Nielsen},\ and\ \citenamefont {Bozhevolnyi}}]{Pors2015}%
  \BibitemOpen
  \bibfield  {author} {\bibinfo {author} {\bibfnamefont {A.}~\bibnamefont
  {Pors}}, \bibinfo {author} {\bibfnamefont {M.~G.}\ \bibnamefont {Nielsen}},\
  and\ \bibinfo {author} {\bibfnamefont {S.~I.}\ \bibnamefont {Bozhevolnyi}},\
  }\bibfield  {title} {\bibinfo {title} {{Analog computing using reflective
  plasmonic metasurfaces}},\ }\href {https://doi.org/10.1021/nl5047297}
  {\bibfield  {journal} {\bibinfo  {journal} {Nano Lett.}\ }\textbf {\bibinfo
  {volume} {15}},\ \bibinfo {pages} {791} (\bibinfo {year} {2015})},\ \Eprint
  {https://arxiv.org/abs/1609.04672} {1609.04672} \BibitemShut {NoStop}%
\bibitem [{\citenamefont {Zhu}\ \emph {et~al.}(2017)\citenamefont {Zhu},
  \citenamefont {Zhou}, \citenamefont {Lou}, \citenamefont {Ye}, \citenamefont
  {Qiu}, \citenamefont {Ruan},\ and\ \citenamefont {Fan}}]{Zhu2017}%
  \BibitemOpen
  \bibfield  {author} {\bibinfo {author} {\bibfnamefont {T.}~\bibnamefont
  {Zhu}}, \bibinfo {author} {\bibfnamefont {Y.}~\bibnamefont {Zhou}}, \bibinfo
  {author} {\bibfnamefont {Y.}~\bibnamefont {Lou}}, \bibinfo {author}
  {\bibfnamefont {H.}~\bibnamefont {Ye}}, \bibinfo {author} {\bibfnamefont
  {M.}~\bibnamefont {Qiu}}, \bibinfo {author} {\bibfnamefont {Z.}~\bibnamefont
  {Ruan}},\ and\ \bibinfo {author} {\bibfnamefont {S.}~\bibnamefont {Fan}},\
  }\bibfield  {title} {\bibinfo {title} {{Plasmonic computing of spatial
  differentiation}},\ }\href {https://doi.org/10.1038/ncomms15391} {\bibfield
  {journal} {\bibinfo  {journal} {Nat. Commun.}\ }\textbf {\bibinfo {volume}
  {8}},\ \bibinfo {pages} {1} (\bibinfo {year} {2017})}\BibitemShut {NoStop}%
\bibitem [{\citenamefont {Kwon}\ \emph {et~al.}(2018)\citenamefont {Kwon},
  \citenamefont {Sounas}, \citenamefont {Cordaro}, \citenamefont {Polman},\
  and\ \citenamefont {Al{\`{u}}}}]{Kwon2018}%
  \BibitemOpen
  \bibfield  {author} {\bibinfo {author} {\bibfnamefont {H.}~\bibnamefont
  {Kwon}}, \bibinfo {author} {\bibfnamefont {D.}~\bibnamefont {Sounas}},
  \bibinfo {author} {\bibfnamefont {A.}~\bibnamefont {Cordaro}}, \bibinfo
  {author} {\bibfnamefont {A.}~\bibnamefont {Polman}},\ and\ \bibinfo {author}
  {\bibfnamefont {A.}~\bibnamefont {Al{\`{u}}}},\ }\bibfield  {title} {\bibinfo
  {title} {{Nonlocal Metasurfaces for Optical Signal Processing}},\ }\href
  {https://doi.org/10.1103/PhysRevLett.121.173004} {\bibfield  {journal}
  {\bibinfo  {journal} {Phys. Rev. Lett.}\ }\textbf {\bibinfo {volume} {121}},\
  \bibinfo {pages} {173004} (\bibinfo {year} {2018})}\BibitemShut {NoStop}%
\bibitem [{\citenamefont {Estakhri}\ \emph {et~al.}(2019)\citenamefont
  {Estakhri}, \citenamefont {Edwards},\ and\ \citenamefont
  {Engheta}}]{Estakhri2019}%
  \BibitemOpen
  \bibfield  {author} {\bibinfo {author} {\bibfnamefont {N.~M.}\ \bibnamefont
  {Estakhri}}, \bibinfo {author} {\bibfnamefont {B.}~\bibnamefont {Edwards}},\
  and\ \bibinfo {author} {\bibfnamefont {N.}~\bibnamefont {Engheta}},\
  }\bibfield  {title} {\bibinfo {title} {{Inverse-designed metastructures that
  solve equations}},\ }\href {https://doi.org/10.1126/science.aaw2498}
  {\bibfield  {journal} {\bibinfo  {journal} {Science}\ }\textbf {\bibinfo
  {volume} {363}},\ \bibinfo {pages} {1333} (\bibinfo {year}
  {2019})}\BibitemShut {NoStop}%
\bibitem [{\citenamefont {Purcell}\ and\ \citenamefont
  {Pennypacker}(1973)}]{Purcell1973}%
  \BibitemOpen
  \bibfield  {author} {\bibinfo {author} {\bibfnamefont {E.~M.}\ \bibnamefont
  {Purcell}}\ and\ \bibinfo {author} {\bibfnamefont {C.~R.}\ \bibnamefont
  {Pennypacker}},\ }\bibfield  {title} {\bibinfo {title} {{Scattering and
  Absorption of Light by Nonspherical Dielectric Grains}},\ }\href
  {https://doi.org/10.1086/152538} {\bibfield  {journal} {\bibinfo  {journal}
  {Astrophys. J.}\ }\textbf {\bibinfo {volume} {186}},\ \bibinfo {pages} {705}
  (\bibinfo {year} {1973})}\BibitemShut {NoStop}%
\bibitem [{\citenamefont {Draine}\ and\ \citenamefont
  {Flatau}(1994)}]{Draine1994}%
  \BibitemOpen
  \bibfield  {author} {\bibinfo {author} {\bibfnamefont {B.~T.}\ \bibnamefont
  {Draine}}\ and\ \bibinfo {author} {\bibfnamefont {P.~J.}\ \bibnamefont
  {Flatau}},\ }\bibfield  {title} {\bibinfo {title} {{Discrete-dipole
  approximation for scattering calculations}},\ }\href
  {https://doi.org/10.1364/JOSAA.11.001491} {\bibfield  {journal} {\bibinfo
  {journal} {J. Opt. Soc. Am. A}\ }\textbf {\bibinfo {volume} {11}},\ \bibinfo
  {pages} {1491} (\bibinfo {year} {1994})}\BibitemShut {NoStop}%
\bibitem [{\citenamefont {Strang}(1994)}]{strang_wavelets_1994}%
  \BibitemOpen
  \bibfield  {author} {\bibinfo {author} {\bibfnamefont {G.}~\bibnamefont
  {Strang}},\ }\bibfield  {title} {\bibinfo {title} {Wavelets},\ }\href
  {https://www.jstor.org/stable/29775194} {\bibfield  {journal} {\bibinfo
  {journal} {American Scientist}\ }\textbf {\bibinfo {volume} {82}},\ \bibinfo
  {pages} {250} (\bibinfo {year} {1994})},\ \bibinfo {note} {publisher: Sigma
  Xi, The Scientific Research Society}\BibitemShut {NoStop}%
\bibitem [{\citenamefont {Rao}\ \emph {et~al.}(2010)\citenamefont {Rao},
  \citenamefont {Kim},\ and\ \citenamefont {Hwang}}]{rao_nonuniform_2010}%
  \BibitemOpen
  \bibfield  {author} {\bibinfo {author} {\bibfnamefont {K.~R.}\ \bibnamefont
  {Rao}}, \bibinfo {author} {\bibfnamefont {D.~N.}\ \bibnamefont {Kim}},\ and\
  \bibinfo {author} {\bibfnamefont {J.~J.}\ \bibnamefont {Hwang}},\ }\bibfield
  {title} {{\selectlanguage {en}\bibinfo {title} {Nonuniform {DFT}}},\ }in\
  \href {https://doi.org/10.1007/978-1-4020-6629-0_7} {{\selectlanguage
  {en}\emph {\bibinfo {booktitle} {Fast {Fourier} {Transform} - {Algorithms}
  and {Applications}}}}},\ \bibinfo {series and number} {Signals and
  {Communication} {Technology}},\ \bibinfo {editor} {edited by\ \bibinfo
  {editor} {\bibfnamefont {K.}~\bibnamefont {Rao}}, \bibinfo {editor}
  {\bibfnamefont {D.}~\bibnamefont {Kim}},\ and\ \bibinfo {editor}
  {\bibfnamefont {J.-J.}\ \bibnamefont {Hwang}}}\ (\bibinfo  {publisher}
  {Springer Netherlands},\ \bibinfo {address} {Dordrecht},\ \bibinfo {year}
  {2010})\ pp.\ \bibinfo {pages} {195--234}\BibitemShut {NoStop}%
\bibitem [{\citenamefont {Hashemi}\ \emph {et~al.}(2012)\citenamefont
  {Hashemi}, \citenamefont {Qiu}, \citenamefont {McCauley}, \citenamefont
  {Joannopoulos},\ and\ \citenamefont {Johnson}}]{Hashemi2012}%
  \BibitemOpen
  \bibfield  {author} {\bibinfo {author} {\bibfnamefont {H.}~\bibnamefont
  {Hashemi}}, \bibinfo {author} {\bibfnamefont {C.-W.}\ \bibnamefont {Qiu}},
  \bibinfo {author} {\bibfnamefont {A.~P.}\ \bibnamefont {McCauley}}, \bibinfo
  {author} {\bibfnamefont {J.~D.}\ \bibnamefont {Joannopoulos}},\ and\ \bibinfo
  {author} {\bibfnamefont {S.~G.}\ \bibnamefont {Johnson}},\ }\bibfield
  {title} {\bibinfo {title} {{Diameter-bandwidth product limitation of
  isolated-object cloaking}},\ }\href
  {https://doi.org/10.1103/PhysRevA.86.013804} {\bibfield  {journal} {\bibinfo
  {journal} {Phys. Rev. A}\ }\textbf {\bibinfo {volume} {86}},\ \bibinfo
  {pages} {013804} (\bibinfo {year} {2012})}\BibitemShut {NoStop}%
\bibitem [{\citenamefont {Yang}\ \emph {et~al.}(2015)\citenamefont {Yang},
  \citenamefont {Antosiewicz}, \citenamefont {Verre}, \citenamefont {{Garcia De
  Abajo}}, \citenamefont {Apell},\ and\ \citenamefont {Kall}}]{Yang2015}%
  \BibitemOpen
  \bibfield  {author} {\bibinfo {author} {\bibfnamefont {Z.~J.}\ \bibnamefont
  {Yang}}, \bibinfo {author} {\bibfnamefont {T.~J.}\ \bibnamefont
  {Antosiewicz}}, \bibinfo {author} {\bibfnamefont {R.}~\bibnamefont {Verre}},
  \bibinfo {author} {\bibfnamefont {F.~J.}\ \bibnamefont {{Garcia De Abajo}}},
  \bibinfo {author} {\bibfnamefont {S.~P.}\ \bibnamefont {Apell}},\ and\
  \bibinfo {author} {\bibfnamefont {M.}~\bibnamefont {Kall}},\ }\bibfield
  {title} {\bibinfo {title} {{Ultimate Limit of Light Extinction by
  Nanophotonic Structures}},\ }\href
  {https://doi.org/10.1021/acs.nanolett.5b03512} {\bibfield  {journal}
  {\bibinfo  {journal} {Nano Lett.}\ }\textbf {\bibinfo {volume} {15}},\
  \bibinfo {pages} {7633} (\bibinfo {year} {2015})}\BibitemShut {NoStop}%
\bibitem [{\citenamefont {Ganapati}\ \emph {et~al.}(2014)\citenamefont
  {Ganapati}, \citenamefont {Miller},\ and\ \citenamefont
  {Yablonovitch}}]{Ganapati2014}%
  \BibitemOpen
  \bibfield  {author} {\bibinfo {author} {\bibfnamefont {V.}~\bibnamefont
  {Ganapati}}, \bibinfo {author} {\bibfnamefont {O.~D.}\ \bibnamefont
  {Miller}},\ and\ \bibinfo {author} {\bibfnamefont {E.}~\bibnamefont
  {Yablonovitch}},\ }\bibfield  {title} {\bibinfo {title} {Light trapping
  textures designed by electromagnetic optimization for subwavelength thick
  solar cells},\ }\href {https://doi.org/10.1109/JPHOTOV.2013.2280340}
  {\bibfield  {journal} {\bibinfo  {journal} {IEEE Journal of Photovoltaics}\
  }\textbf {\bibinfo {volume} {4}},\ \bibinfo {pages} {175} (\bibinfo {year}
  {2014})},\ \Eprint {https://arxiv.org/abs/1307.5465} {1307.5465} \BibitemShut
  {NoStop}%
\bibitem [{\citenamefont {Buddhiraju}\ and\ \citenamefont
  {Fan}(2017)}]{Buddhiraju2017}%
  \BibitemOpen
  \bibfield  {author} {\bibinfo {author} {\bibfnamefont {S.}~\bibnamefont
  {Buddhiraju}}\ and\ \bibinfo {author} {\bibfnamefont {S.}~\bibnamefont
  {Fan}},\ }\bibfield  {title} {\bibinfo {title} {Theory of solar cell light
  trapping through a nonequilibrium green's function formulation of maxwell's
  equations},\ }\href {https://doi.org/10.1103/PhysRevB.96.035304} {\bibfield
  {journal} {\bibinfo  {journal} {Physical Review B}\ }\textbf {\bibinfo
  {volume} {96}},\ \bibinfo {pages} {035304} (\bibinfo {year}
  {2017})}\BibitemShut {NoStop}%
\bibitem [{\citenamefont {Benzaouia}\ \emph {et~al.}(2019)\citenamefont
  {Benzaouia}, \citenamefont {Tokic}, \citenamefont {Miller}, \citenamefont
  {Yue},\ and\ \citenamefont {Johnson}}]{Benzaouia2019}%
  \BibitemOpen
  \bibfield  {author} {\bibinfo {author} {\bibfnamefont {M.}~\bibnamefont
  {Benzaouia}}, \bibinfo {author} {\bibfnamefont {G.}~\bibnamefont {Tokic}},
  \bibinfo {author} {\bibfnamefont {O.~D.}\ \bibnamefont {Miller}}, \bibinfo
  {author} {\bibfnamefont {D.~K.~P.}\ \bibnamefont {Yue}},\ and\ \bibinfo
  {author} {\bibfnamefont {S.~G.}\ \bibnamefont {Johnson}},\ }\bibfield
  {title} {\bibinfo {title} {From solar cells to ocean buoys: Wide-bandwidth
  limits to absorption by metaparticle arrays},\ }\href
  {https://doi.org/10.1103/PhysRevApplied.11.034033} {\bibfield  {journal}
  {\bibinfo  {journal} {Physical Review Applied}\ }\textbf {\bibinfo {volume}
  {11}},\ \bibinfo {pages} {034033} (\bibinfo {year} {2019})},\ \Eprint
  {https://arxiv.org/abs/1804.00600} {1804.00600} \BibitemShut {NoStop}%
\bibitem [{\citenamefont {Weiner}(2011)}]{Weiner2011}%
  \BibitemOpen
  \bibfield  {author} {\bibinfo {author} {\bibfnamefont {A.}~\bibnamefont
  {Weiner}},\ }\href@noop {} {\emph {\bibinfo {title} {{Ultrafast optics}}}},\
  Vol.~\bibinfo {volume} {72}\ (\bibinfo  {publisher} {John Wiley {\&} Sons},\
  \bibinfo {year} {2011})\BibitemShut {NoStop}%
\bibitem [{\citenamefont {Agrawal}(2013)}]{Agrawal2013}%
  \BibitemOpen
  \bibfield  {author} {\bibinfo {author} {\bibfnamefont {G.~P.}\ \bibnamefont
  {Agrawal}},\ }\href@noop {} {\emph {\bibinfo {title} {{Nonlinear Fiber
  Optics}}}},\ \bibinfo {edition} {5th}\ ed.\ (\bibinfo  {publisher} {Academic
  Press},\ \bibinfo {year} {2013})\BibitemShut {NoStop}%
\bibitem [{\citenamefont {Wang}\ \emph {et~al.}(2008)\citenamefont {Wang},
  \citenamefont {Rozhin}, \citenamefont {Scardaci}, \citenamefont {Sun},
  \citenamefont {Hennrich}, \citenamefont {White}, \citenamefont {Milne},\ and\
  \citenamefont {Ferrari}}]{Wang2008}%
  \BibitemOpen
  \bibfield  {author} {\bibinfo {author} {\bibfnamefont {F.}~\bibnamefont
  {Wang}}, \bibinfo {author} {\bibfnamefont {A.~G.}\ \bibnamefont {Rozhin}},
  \bibinfo {author} {\bibfnamefont {V.}~\bibnamefont {Scardaci}}, \bibinfo
  {author} {\bibfnamefont {Z.}~\bibnamefont {Sun}}, \bibinfo {author}
  {\bibfnamefont {F.}~\bibnamefont {Hennrich}}, \bibinfo {author}
  {\bibfnamefont {I.~H.}\ \bibnamefont {White}}, \bibinfo {author}
  {\bibfnamefont {W.~I.}\ \bibnamefont {Milne}},\ and\ \bibinfo {author}
  {\bibfnamefont {A.~C.}\ \bibnamefont {Ferrari}},\ }\bibfield  {title}
  {\bibinfo {title} {{Wideband-tuneable, nanotube mode-locked, fibre laser}},\
  }\href {https://doi.org/10.1038/nnano.2008.312} {\bibfield  {journal}
  {\bibinfo  {journal} {Nat. Nanotechnol.}\ }\textbf {\bibinfo {volume} {3}},\
  \bibinfo {pages} {738} (\bibinfo {year} {2008})}\BibitemShut {NoStop}%
\bibitem [{\citenamefont {Rokhlin}(1985)}]{rokhlin_rapid_1985}%
  \BibitemOpen
  \bibfield  {author} {\bibinfo {author} {\bibfnamefont {V.}~\bibnamefont
  {Rokhlin}},\ }\bibfield  {title} {{\selectlanguage {en}\bibinfo {title}
  {Rapid solution of integral equations of classical potential theory}},\
  }\href {https://doi.org/10.1016/0021-9991(85)90002-6} {\bibfield  {journal}
  {\bibinfo  {journal} {Journal of Computational Physics}\ }\textbf {\bibinfo
  {volume} {60}},\ \bibinfo {pages} {187} (\bibinfo {year} {1985})}\BibitemShut
  {NoStop}%
\bibitem [{\citenamefont {Harrington}(1993)}]{harrington_field_1993-1}%
  \BibitemOpen
  \bibfield  {author} {\bibinfo {author} {\bibfnamefont {R.~F.}\ \bibnamefont
  {Harrington}},\ }\href@noop {} {\emph {\bibinfo {title} {Field {Computation}
  by {Moment} {Methods}}}}\ (\bibinfo  {publisher} {Wiley-IEEE Press},\
  \bibinfo {year} {1993})\BibitemShut {NoStop}%
\bibitem [{\citenamefont {Johnson}\ and\ \citenamefont
  {Joannopoulos}(2001)}]{johnson_block-iterative_2001}%
  \BibitemOpen
  \bibfield  {author} {\bibinfo {author} {\bibfnamefont {S.~G.}\ \bibnamefont
  {Johnson}}\ and\ \bibinfo {author} {\bibfnamefont {J.~D.}\ \bibnamefont
  {Joannopoulos}},\ }\bibfield  {title} {{\selectlanguage {EN}\bibinfo {title}
  {Block-iterative frequency-domain methods for {Maxwell}’s equations in a
  planewave basis}},\ }\href {https://doi.org/10.1364/OE.8.000173} {\bibfield
  {journal} {\bibinfo  {journal} {Optics Express}\ }\textbf {\bibinfo {volume}
  {8}},\ \bibinfo {pages} {173} (\bibinfo {year} {2001})}\BibitemShut {NoStop}%
\end{thebibliography}%


\begin{thebibliography}{14}%
\makeatletter
\providecommand \@ifxundefined [1]{%
 \@ifx{#1\undefined}
}%
\providecommand \@ifnum [1]{%
 \ifnum #1\expandafter \@firstoftwo
 \else \expandafter \@secondoftwo
 \fi
}%
\providecommand \@ifx [1]{%
 \ifx #1\expandafter \@firstoftwo
 \else \expandafter \@secondoftwo
 \fi
}%
\providecommand \natexlab [1]{#1}%
\providecommand \enquote  [1]{``#1''}%
\providecommand \bibnamefont  [1]{#1}%
\providecommand \bibfnamefont [1]{#1}%
\providecommand \citenamefont [1]{#1}%
\providecommand \href@noop [0]{\@secondoftwo}%
\providecommand \href [0]{\begingroup \@sanitize@url \@href}%
\providecommand \@href[1]{\@@startlink{#1}\@@href}%
\providecommand \@@href[1]{\endgroup#1\@@endlink}%
\providecommand \@sanitize@url [0]{\catcode `\\12\catcode `\$12\catcode
  `\&12\catcode `\#12\catcode `\^12\catcode `\_12\catcode `\%12\relax}%
\providecommand \@@startlink[1]{}%
\providecommand \@@endlink[0]{}%
\providecommand \url  [0]{\begingroup\@sanitize@url \@url }%
\providecommand \@url [1]{\endgroup\@href {#1}{\urlprefix }}%
\providecommand \urlprefix  [0]{URL }%
\providecommand \Eprint [0]{\href }%
\providecommand \doibase [0]{https://doi.org/}%
\providecommand \selectlanguage [0]{\@gobble}%
\providecommand \bibinfo  [0]{\@secondoftwo}%
\providecommand \bibfield  [0]{\@secondoftwo}%
\providecommand \translation [1]{[#1]}%
\providecommand \BibitemOpen [0]{}%
\providecommand \bibitemStop [0]{}%
\providecommand \bibitemNoStop [0]{.\EOS\space}%
\providecommand \EOS [0]{\spacefactor3000\relax}%
\providecommand \BibitemShut  [1]{\csname bibitem#1\endcsname}%
\let\auto@bib@innerbib\@empty
%</preamble>
\bibitem [{\citenamefont {Chew}\ \emph {et~al.}(2008)\citenamefont {Chew},
  \citenamefont {Tong},\ and\ \citenamefont {Hu}}]{Chew2008}%
  \BibitemOpen
  \bibfield  {author} {\bibinfo {author} {\bibfnamefont {W.~C.}\ \bibnamefont
  {Chew}}, \bibinfo {author} {\bibfnamefont {M.~S.}\ \bibnamefont {Tong}},\
  and\ \bibinfo {author} {\bibfnamefont {B.}~\bibnamefont {Hu}},\ }\bibfield
  {title} {\bibinfo {title} {{Integral equation methods for electromagnetic and
  elastic waves}},\ }\href@noop {} {\bibfield  {journal} {\bibinfo  {journal}
  {Synth. Lect. Comput. Electromagn.}\ }\textbf {\bibinfo {volume} {3}},\
  \bibinfo {pages} {1} (\bibinfo {year} {2008})}\BibitemShut {NoStop}%
\bibitem [{\citenamefont {Ben-Tal}\ and\ \citenamefont
  {Teboulle}(1996)}]{ben-tal_hidden_1996}%
  \BibitemOpen
  \bibfield  {author} {\bibinfo {author} {\bibfnamefont {A.}~\bibnamefont
  {Ben-Tal}}\ and\ \bibinfo {author} {\bibfnamefont {M.}~\bibnamefont
  {Teboulle}},\ }\bibfield  {title} {{\selectlanguage {en}\bibinfo {title}
  {Hidden convexity in some nonconvex quadratically constrained quadratic
  programming}},\ }\href {https://doi.org/10.1007/BF02592331} {\bibfield
  {journal} {\bibinfo  {journal} {Mathematical Programming}\ }\textbf {\bibinfo
  {volume} {72}},\ \bibinfo {pages} {51} (\bibinfo {year} {1996})}\BibitemShut
  {NoStop}%
\bibitem [{\citenamefont {Luo}\ \emph {et~al.}(2010)\citenamefont {Luo},
  \citenamefont {Ma}, \citenamefont {So}, \citenamefont {Ye},\ and\
  \citenamefont {Zhang}}]{Luo2010}%
  \BibitemOpen
  \bibfield  {author} {\bibinfo {author} {\bibfnamefont {Z.~Q.}\ \bibnamefont
  {Luo}}, \bibinfo {author} {\bibfnamefont {W.~K.}\ \bibnamefont {Ma}},
  \bibinfo {author} {\bibfnamefont {A.}~\bibnamefont {So}}, \bibinfo {author}
  {\bibfnamefont {Y.}~\bibnamefont {Ye}},\ and\ \bibinfo {author}
  {\bibfnamefont {S.}~\bibnamefont {Zhang}},\ }\bibfield  {title} {\bibinfo
  {title} {{Semidefinite relaxation of quadratic optimization problems}},\
  }\href {https://doi.org/10.1109/MSP.2010.936019} {\bibfield  {journal}
  {\bibinfo  {journal} {IEEE Signal Process. Mag.}\ }\textbf {\bibinfo {volume}
  {27}},\ \bibinfo {pages} {20} (\bibinfo {year} {2010})}\BibitemShut {NoStop}%
\bibitem [{\citenamefont {Park}\ and\ \citenamefont
  {Boyd}(2017)}]{park_general_2017}%
  \BibitemOpen
  \bibfield  {author} {\bibinfo {author} {\bibfnamefont {J.}~\bibnamefont
  {Park}}\ and\ \bibinfo {author} {\bibfnamefont {S.}~\bibnamefont {Boyd}},\
  }\bibfield  {title} {\bibinfo {title} {General {Heuristics} for {Nonconvex}
  {Quadratically} {Constrained} {Quadratic} {Programming}},\ }\href
  {http://arxiv.org/abs/1703.07870} {\bibfield  {journal} {\bibinfo  {journal}
  {arXiv:1703.07870 [math]}\ } (\bibinfo {year} {2017})},\ \bibinfo {note}
  {arXiv: 1703.07870}\BibitemShut {NoStop}%
\bibitem [{\citenamefont {Lytle}\ \emph {et~al.}(2005)\citenamefont {Lytle},
  \citenamefont {Carney}, \citenamefont {Schotland},\ and\ \citenamefont
  {Wolf}}]{Lytle2005}%
  \BibitemOpen
  \bibfield  {author} {\bibinfo {author} {\bibfnamefont {D.~R.}\ \bibnamefont
  {Lytle}}, \bibinfo {author} {\bibfnamefont {P.~S.}\ \bibnamefont {Carney}},
  \bibinfo {author} {\bibfnamefont {J.~C.}\ \bibnamefont {Schotland}},\ and\
  \bibinfo {author} {\bibfnamefont {E.}~\bibnamefont {Wolf}},\ }\bibfield
  {title} {\bibinfo {title} {{Generalized optical theorem for reflection,
  transmission, and extinction of power for electromagnetic fields}},\ }\href
  {https://doi.org/10.1103/PhysRevE.71.056610} {\bibfield  {journal} {\bibinfo
  {journal} {Phys. Rev. E}\ }\textbf {\bibinfo {volume} {71}},\ \bibinfo
  {pages} {056610} (\bibinfo {year} {2005})}\BibitemShut {NoStop}%
\bibitem [{\citenamefont {Newton}(1976)}]{Newton1976}%
  \BibitemOpen
  \bibfield  {author} {\bibinfo {author} {\bibfnamefont {R.~G.}\ \bibnamefont
  {Newton}},\ }\bibfield  {title} {\bibinfo {title} {{Optical theorem and
  beyond}},\ }\href {https://doi.org/10.1119/1.10324} {\bibfield  {journal}
  {\bibinfo  {journal} {Am. J. Phys.}\ }\textbf {\bibinfo {volume} {44}},\
  \bibinfo {pages} {639} (\bibinfo {year} {1976})}\BibitemShut {NoStop}%
\bibitem [{\citenamefont {Jackson}(1999)}]{Jackson1999}%
  \BibitemOpen
  \bibfield  {author} {\bibinfo {author} {\bibfnamefont {J.~D.}\ \bibnamefont
  {Jackson}},\ }\href@noop {} {\emph {\bibinfo {title} {{Classical
  Electrodynamics, 3rd Ed.}}}}\ (\bibinfo  {publisher} {John Wiley {\&} Sons},\
  \bibinfo {year} {1999})\BibitemShut {NoStop}%
\bibitem [{\citenamefont {Hashemi}\ \emph {et~al.}(2012)\citenamefont
  {Hashemi}, \citenamefont {Qiu}, \citenamefont {McCauley}, \citenamefont
  {Joannopoulos},\ and\ \citenamefont {Johnson}}]{Hashemi2012}%
  \BibitemOpen
  \bibfield  {author} {\bibinfo {author} {\bibfnamefont {H.}~\bibnamefont
  {Hashemi}}, \bibinfo {author} {\bibfnamefont {C.-W.}\ \bibnamefont {Qiu}},
  \bibinfo {author} {\bibfnamefont {A.~P.}\ \bibnamefont {McCauley}}, \bibinfo
  {author} {\bibfnamefont {J.~D.}\ \bibnamefont {Joannopoulos}},\ and\ \bibinfo
  {author} {\bibfnamefont {S.~G.}\ \bibnamefont {Johnson}},\ }\bibfield
  {title} {\bibinfo {title} {{Diameter-bandwidth product limitation of
  isolated-object cloaking}},\ }\href
  {https://doi.org/10.1103/PhysRevA.86.013804} {\bibfield  {journal} {\bibinfo
  {journal} {Phys. Rev. A}\ }\textbf {\bibinfo {volume} {86}},\ \bibinfo
  {pages} {013804} (\bibinfo {year} {2012})}\BibitemShut {NoStop}%
\bibitem [{\citenamefont {Shim}\ \emph {et~al.}(2019)\citenamefont {Shim},
  \citenamefont {Fan}, \citenamefont {Johnson},\ and\ \citenamefont
  {Miller}}]{Shim2019}%
  \BibitemOpen
  \bibfield  {author} {\bibinfo {author} {\bibfnamefont {H.}~\bibnamefont
  {Shim}}, \bibinfo {author} {\bibfnamefont {L.}~\bibnamefont {Fan}}, \bibinfo
  {author} {\bibfnamefont {S.~G.}\ \bibnamefont {Johnson}},\ and\ \bibinfo
  {author} {\bibfnamefont {O.~D.}\ \bibnamefont {Miller}},\ }\bibfield  {title}
  {\bibinfo {title} {{Fundamental Limits to Near-Field Optical Response over
  Any Bandwidth}},\ }\href {https://doi.org/10.1103/PhysRevX.9.011043}
  {\bibfield  {journal} {\bibinfo  {journal} {Phys. Rev. X}\ }\textbf {\bibinfo
  {volume} {9}},\ \bibinfo {pages} {11043} (\bibinfo {year}
  {2019})}\BibitemShut {NoStop}%
\bibitem [{\citenamefont {Landau}\ and\ \citenamefont
  {Lifshitz}(1960)}]{Landau1960}%
  \BibitemOpen
  \bibfield  {author} {\bibinfo {author} {\bibfnamefont {L.~D.}\ \bibnamefont
  {Landau}}\ and\ \bibinfo {author} {\bibfnamefont {E.~M.}\ \bibnamefont
  {Lifshitz}},\ }\href@noop {} {\emph {\bibinfo {title} {{Electrodynamics of
  Continuous Media}}}}\ (\bibinfo  {publisher} {Pergamon Press},\ \bibinfo
  {year} {1960})\BibitemShut {NoStop}%
\bibitem [{\citenamefont {Nussenzveig}(1972)}]{Nussenzveig1972}%
  \BibitemOpen
  \bibfield  {author} {\bibinfo {author} {\bibfnamefont {H.~M.}\ \bibnamefont
  {Nussenzveig}},\ }\href@noop {} {\emph {\bibinfo {title} {{Causality and
  Dispersion Relations}}}}\ (\bibinfo  {publisher} {Academic Press},\ \bibinfo
  {address} {New York, NY},\ \bibinfo {year} {1972})\BibitemShut {NoStop}%
\bibitem [{\citenamefont {Zemanian}(1970)}]{zemanian_hilbert_1970}%
  \BibitemOpen
  \bibfield  {author} {\bibinfo {author} {\bibfnamefont {A.~H.}\ \bibnamefont
  {Zemanian}},\ }\bibfield  {title} {\bibinfo {title} {The {Hilbert} port},\
  }\href@noop {} {\bibfield  {journal} {\bibinfo  {journal} {SIAM Journal on
  Applied Mathematics}\ }\textbf {\bibinfo {volume} {18}},\ \bibinfo {pages}
  {98} (\bibinfo {year} {1970})},\ \bibinfo {note} {publisher:
  SIAM}\BibitemShut {NoStop}%
\bibitem [{\citenamefont {Zemanian}(1995)}]{zemanian_realizability_1995}%
  \BibitemOpen
  \bibfield  {author} {\bibinfo {author} {\bibfnamefont {A.~H.}\ \bibnamefont
  {Zemanian}},\ }\href@noop {} {\emph {\bibinfo {title} {Realizability theory
  for continuous linear systems}}}\ (\bibinfo  {publisher} {Courier
  Corporation},\ \bibinfo {year} {1995})\BibitemShut {NoStop}%
\bibitem [{\citenamefont {Welters}\ \emph {et~al.}(2014)\citenamefont
  {Welters}, \citenamefont {Avniel},\ and\ \citenamefont
  {Johnson}}]{welters_speed--light_2014}%
  \BibitemOpen
  \bibfield  {author} {\bibinfo {author} {\bibfnamefont {A.}~\bibnamefont
  {Welters}}, \bibinfo {author} {\bibfnamefont {Y.}~\bibnamefont {Avniel}},\
  and\ \bibinfo {author} {\bibfnamefont {S.~G.}\ \bibnamefont {Johnson}},\
  }\bibfield  {title} {\bibinfo {title} {Speed-of-light limitations in passive
  linear media},\ }\bibfield  {journal} {\bibinfo  {journal} {Physical Review
  A}\ }\textbf {\bibinfo {volume} {90}},\ \href
  {https://doi.org/10.1103/PhysRevA.90.023847} {10.1103/PhysRevA.90.023847}
  (\bibinfo {year} {2014})\BibitemShut {NoStop}%
\end{thebibliography}%

\end{document}

\title{Supplementary Material: Computational bounds to light--matter interactions via local conservation laws}

\author{Zeyu Kuang}
\affiliation{Department of Applied Physics and Energy Sciences Institute, Yale University, New Haven, Connecticut 06511, USA}
\author{Owen D. Miller}
\affiliation{Department of Applied Physics and Energy Sciences Institute, Yale University, New Haven, Connecticut 06511, USA}

\date{\today}

\maketitle

\tableofcontents

\hl{
\section{Generalized local conservation laws}
In the main text, we \hl{use physical intuition, from the complex Poynting theorem, to argue that} the following local energy conservation laws hold for any geometry in a designable region $V$:
\begin{align}
-\frac{\omega^*}{2}\int_V \phi^\dagger(x) \DD(x) \psi_{\rm inc}(x)\td x  = \frac{\omega^*}{2}\int_V \phi^\dagger(x) \DD(x) \td x \left[\int_V \Gamma_0(x,x')\phi(x')\td x' - \chi^{-1}\phi(x)\right], \label{eq:loc_consv_x}
\end{align}
\hl{where $\phi$ represents the polarization fields, $\Gamma_0$ is the background Green's function, $\psi_{\rm inc}$ is the incident field, and $\DD$ is a 6$\times$6 tensor field that acts as a weighting function over space and polarization. We have allowed the frequency to possibly be complex and taken its conjugate, which proves useful for transforming bandwidth-averaged scattering problems into single-complex-frequency problems.}

In this section, we \hl{provide a rigorous and direct, but perhaps less physically intuitive, derivation of \eqref{loc_consv_x}. We start assuming only Maxwell's equations, and as a first step perform a standard reformulation into \emph{volume integral equations}~\cite{Chew2008}, various forms of which are sometimes referred to as the Lippmann--Schwinger equation. We start by simply equating the total field at any point, $\psi(x)$, to the sum of the} incident field $\psi_{\rm inc}(x)$ and scattered field $\psi_{\rm scat}(x)$:
\begin{equation}
\psi(x) = \psi_{\rm inc}(x) + \psi_{\rm scat}(x).
\label{eq:VIE0}
\end{equation}
\hl{Scattered fields, by definition, are the fields radiated by the polarization currents $\phi(x)$, across the scatterer, that are induced by the presence of the scatterer:} $\psi_{\rm scat}(x) = \int_V \Gamma_0(x,x')\phi(x')\td x'$, where $\Gamma_0(x,x')$ is again the background Green's function. \hl{We want to write \eqref{VIE0} solely in terms of the polarization fields; to do so, we note that at any point in the scatterer the polarization field is given by $\phi(x) = \chi \psi(x)$, where $\chi$ is the material susceptibility tensor, which can be inverted:} $\psi(x) = \chi^{-1}\phi(x)$. Substituting both $\psi(x)$ and $\psi_{\rm scat}(x)$ as a function of $\phi(x)$ in \eqref{VIE0} and rearranging the terms leads to:
\begin{equation}
-\psiinc(x) = \int_V \Gamma_0(x,x')\phi(x')\td x' - \chi^{-1}\phi(x),\quad\quad x\in V_{\rm material}.
\label{eq:VIE1}
\end{equation}
In our six-vector notation, the electric and magnetic fields are stacked together with six polarization components in total, and  \eqref{VIE1} holds for each \hl{polarization index $i$}: 
\begin{equation}
-\psi_{\text{inc}, i}(x) = \left[\int_V \Gamma_0(x,x')\phi(x')\td x'\right]_i - \left[\chi^{-1}\phi(x)\right]_i,\quad\quad x\in V_{\rm material}.
\label{eq:VIE2}
\end{equation}
The index $i = 1, 2, ..., 6$ runs through the six polarization components.
Next, we correlate every polarization component by multiplying \eqref{VIE2} \hl{by} the $j$-th polarization component $\phi^*_j(x)$ at the \textit{same} point in the material:
\begin{equation}
-\phi^*_j(x) \psi_{\text{inc}, i}(x) = \phi^*_j(x) \left[\int_V \Gamma_0(x,x')\phi(x')\td x'\right]_i - \phi^*_j(x) \left[\chi^{-1}\phi(x)\right]_i,\quad\quad x\in V_{\rm material}.
\label{eq:VIE3}
\end{equation}
\hl{This multiplication by $\phi_j^*(x)$ to create the conservation law of \eqref{VIE3} serves a very important purpose. \Eqref{VIE3} is now not only valid over the entire material region, $V_{\rm material}$, it can also be extended throughout the background-material region that comprises the remainder of the designable-region domain $V$, keeping $\chi^{-1}$ as a constant tensor. This is trivially true because $\phi_j^*(x)$ is zero outside of $V_{\rm material}$. By contrast, in the original volume-integral equation of \eqref{VIE0}, $\phi(x)$ for $x \in V_{\rm material}$ is only present in one of the three terms; that equation, therefore, cannot be extended outside of the material domain. Thus, we have now this crucial property of ``domain-obliviousness:'' we can extend \eqref{VIE3} to any point in the designable region $V$:
\begin{equation}
-\phi^*_j(x) \psi_{\text{inc}, i}(x) = \phi^*_j(x) \left[\int_V \Gamma_0(x,x')\phi(x')\td x'\right]_i - \phi^*_j(x) \left[\chi^{-1}\phi(x)\right]_i,\quad\quad x\in V.
\label{eq:VIE4}
\end{equation}
\Eqref{VIE4} represents infinitely many constraints over the domain of the designable region. Any one of the constraints is a pointwise conservation law representing a generalized complex Poynting theorem.

In practice, we can only impose a finite set of constraints. The optimal constraints to use are not necessarily simply a subset of the pointwise constraints. Instead, we can take weighted averages of \eqref{VIE4} over polarization and space to form a new space of constraints out of all possible linear combinations. To prepare for such an average, we first multiply \eqref{VIE4} by a space- and polarization-dependent coefficient $d_{ij}(x)$:
\begin{equation}
    -\phi^*_j(x) d_{ij}(x) \psi_{\text{inc}, i}(x) = \phi^*_j(x) d_{ji}(x) \left[\int_V \Gamma_0(x,x')\phi(x')\td x'\right]_i - \phi^*_j(x) d_{ji}(x) \left[\chi^{-1}\phi(x)\right]_i,\quad\quad x\in V.
\label{eq:VIE5}
\end{equation}
Now, when we sum \eqref{VIE5} over polarizations and integrate over $V$, we have:
\begin{align}
-\int_V \phi^\dagger(x) \DD(x) \psi_{\rm inc}(x)\td x  = \int_V \phi^\dagger(x) \DD(x) \td x \int_V \Gamma_0(x,x')\phi(x')\td x' - \int_V \phi^\dagger(x) \DD(x) \chi^{-1}\phi(x)\td x. \label{eq:VIE6}
\end{align}
Now we can identify an infinite set of constraints through the infinite set of $\DD$ tensors that are possible. If we multiply \eqref{VIE6} by $\omega^*/2$, we have precisely the expression of \eqref{loc_consv_x}. In matrix notation (i.e., assuming any standard discretization), they are equivalent to
\begin{align}
     \frac{\omega^*}{2} \phi^\dagger \DD\GO \phi - \frac{\omega^*}{2} \phi^\dagger \DD \chi^{-1} \phi = -\frac{\omega^*}{2} \phi^\dagger\DD\psi_{\rm inc},
    \label{eq:optthm}
\end{align}
which is Eq.~(3) of the main text.
}}

\section{Semidefinite relaxation of the QCQP problem}
\label{sec:SDR}
In the main text, we show that one can formulate the bound problem with a quadratic-form objective and the conservation-law constraints:
\begin{equation}
    \begin{aligned}
        & \underset{\phi}{\text{max.}}
        & & f(\phi) = \phi^\dagger \mathbb{A} \phi +  \Re\left(\beta^\dagger \phi\right)+c \\
        & \text{s.t.}
        & &  \phi^\dagger \Re\left\{\DD_j\omega^*(\xi+\Gamma_0)\right\} \phi = -\Re \left(\omega^*\phi^\dagger\DD_j\psiinc\right),
    \end{aligned}
    \label{eq:opt_prob}
\end{equation}
\hl{where, for simplicity, we have introduced} \hl{a new variable $\xi = -\chi^{-1}$. This type of problem with quadratic objective and quadratic constraints is well studied in the optimization literature~\cite{ben-tal_hidden_1996,Luo2010,park_general_2017}.}
In this section we \hl{describe how it is} translated to a semidefinite program using standard techniques: \hl{each step below is also clearly explained in \citeasnoun{Luo2010}}. The first step is to \emph{homogenize} the quadratic forms on \eqref{opt_prob}, which means introducing an additional variable in order to have purely quadratic and scalar terms without any linear term. To do this, in the objective function we introduce a complex-valued scalar variable \hl{$s$} into the linear term:
\begin{align}
    f(\phi) = \phi^\dagger \mathbb{A} \phi +  \Re\left(\hl{s}^\dagger \beta^\dagger \phi\right)+c.
\end{align}
The key advantage of introducing this variable is that now one can write $f$ as a homogeneous quadratic form:
\begin{align}
    f\left(
        \begin{bmatrix}
            \phi \\
            \hl{s}
        \end{bmatrix}
    \right)
    = 
    \begin{pmatrix}
        \phi \\
        \hl{s}
    \end{pmatrix}^\dagger 
    \begin{pmatrix}
        \mathbb{A} & \frac{1}{2}\beta \\
        \frac{1}{2}\beta^\dagger & 0
    \end{pmatrix}
    \begin{pmatrix}
        \phi \\
        \hl{s}
    \end{pmatrix}
    + c.
\end{align}
We can do this for each of the $j$ constraints as well, introducing \hl{the} dummy variable \hl{s} for each constraint, which then takes the form:
\begin{align}
    \begin{pmatrix}
        \phi \\
        \hl{s}
    \end{pmatrix}^\dagger 
    \begin{pmatrix}
        \Re\left\{\DD_j\omega^*(\xi+\Gamma_0)\right\} & \frac{1}{2}\omega^* \DD_j \psiinc \\
        \frac{1}{2}\omega \psiinc^\dagger \DD_j^\dagger & 0
    \end{pmatrix}
    \begin{pmatrix}
        \phi \\
        \hl{s}
    \end{pmatrix}
    = 0.
\end{align}
One cannot allow \hl{$s$} to take arbitrary values or else it will modify the initial problem. Instead, \hl{it} should be required to have modulus one, i.e., $|\hl{s}|^2 = 1$, which is itself a quadratic form in the degrees of freedom \hl{$\phi$ and $s$}. Finally, we can lump all degrees of freedom into a single vector $v$:
\begin{align}
    v = 
    \begin{pmatrix}
        \phi \\
        \hl{s}
    \end{pmatrix}.
\end{align}
With this notation, the objective, the $N$ conservation-law constraints, and \hl{the modulus constraint of $s$ are all written} in the form 
\begin{align}
    v^\dagger \mathbb{F} v.
\end{align}
The way to optimize over such quadratic forms is to ``lift'' them to a higher-dimensional space where they become linear forms. The first step is to use the trace operator to rewrite the quadratic form:
\begin{align}
    v^\dagger \mathbb{F} v = \Tr{\mathbb{F} v v^\dagger}.
\end{align}
Then one defines a \hl{rank-one} \emph{matrix variable} $\mathbb{X}$ given by $vv^\dagger$, in which case we now have a linear form:
\begin{align}
    \Tr{\mathbb{F} vv^\dagger} = \Tr{\mathbb{F} \mathbb{X}}.
\end{align}
One cannot optimize arbitrarily over $\mathbb{X}$ and have an equivalent problem; one must additionally impose constraints that $\mathbb{X}$ be a rank-one, positive-definite matrix. The rank-one constraint is nonconvex; the ``relaxation'' in semidefinite relaxation (SDR) refers to dropping this rank-one constraint. Once that constraint has been removed, one is left with a linear objective function (in $\mathbb{X}$), and $2N+1$ linear constraints, over the space of positive-definite matrices. The transformation to a semidefinite program is complete.

\hl{The semidefinite relaxation mentioned above does not introduce any actual relaxation if there is only one (global) constraint in \hl{the} optimization problem~\cite{Luo2010}. Furthermore, even though a certain degree of relaxation may be triggered by  additional (local) constraints, \hl{it is straightforward to show that the additional constraints can \emph{only} tighten the bound.} This can also be seen in our examples for both absorption cross-section (Fig. 1(b) in the main text) and broadband extinction (Fig. 3 in the main text), where the bounds are always monotonically decreasing with the additional local constraints.
}

\section{Algorithm: Maximally violated local constraints} \label{sec:AM}
In this section we derive the optimal new $\DD$ matrix, and corresponding conservation-law constraint, that should be added to a given set of constraints by our principle of maximum violation. The conservation-law constraints as given in the main text are of the form
\begin{align}
    \phi^\dagger \Re\left\{\DD_j\omega^*(\xi+\Gamma_0)\right\} \phi = -\Re \left(\omega^*\phi^\dagger\DD_j\psiinc\right),
    \label{eq:loc_law}
\end{align}
where $j$ runs from 1 to $N$, where $N$ is the current number of constraints that have been imposed. For simplicity, we \hl{have introduced} $\xi=-\chi^{-1}$. The key remaining question, then, is how to select the $(N+1)^{\rm th}$ constraint? From the first $N$ constraints, one can identify a potentially optimal polarization current $\phi_{\rm opt}$ as the first singular vector of the optimal matrix solution of the SDP (as discussed in \secref{SDR}). Given this polarization current, then, a sensible approach to selecting a new constraint is to identify the constraint whose residual is largest when evaluated for polarization current $\phi_{\rm opt}$. In other words, we want the $\DD_{N+1}$ that maximizes the quantity
\begin{equation}
    \begin{aligned}
        & \underset{\DD}{\text{maximize}}
        & & \left| \Re\left\{ \phi_{\rm opt}^\dagger \DD\omega^*(\xi+\Gamma_0) \phi_{\rm opt} + \omega^*\phi_{\rm opt}^\dagger\DD\psiinc \right\} \right|.
    \end{aligned}
\end{equation}
By the cyclic property of the matrix trace, we can rewrite this expression as
\begin{align}
    \Re\Tr{ \DD\left[ \omega^*(\xi+\Gamma_0) \phi_{\rm opt} \phi_{\rm opt}^\dagger + \omega^* \psiinc \phi_{\rm opt}^\dagger \right] },
\end{align}
where we dropped the absolute value since any optimal negative value can be reversed through $\DD \rightarrow -\DD$. Let us denote the matrix in square brackets as $\mathbb{C}$. Expanding the real (Hermitian) part, we have
\begin{align}
    \frac{1}{2} \left[ \Tr {\DD\mathbb{C} } + \Tr{\DD^\dagger \mathbb{C}^\dagger} \right].
\end{align}
Clearly one can maximize the residual by allowing the norm of $\DD$ to be arbitrarily large, but that would not give insight into which spatial pattern $\DD$ should take. As a normalization we can take the Frobenius norm of $\DD$ to be 1, i.e. $\Tr{\DD^\dagger \DD} = 1$. Then, straightforward variational calculus yields an optimal $\DD$ matrix given by $\DD = \mathbb{C}^\dagger$; since $\DD$ must be (spatially) diagonal, we take $\DD$ to comprise the diagonal elements of $\mathbb{C}^\dagger$:
\begin{align}
    \DD_{N+1} &= \diag\left[\mathbb{C}^\dagger\right] \nonumber \\
              &= \omega \diag\left[\phi_{\rm opt} \phi_{\rm opt}^\dagger \left(\xi+\Gamma_0\right)^\dagger + \phi_{\rm opt}\psiinc \right],
              \label{eq:D_iter}
\end{align}
where now ``$\diag$'' strips its matrix argument of all elements except along the (spatial) diagonal, as in the main text. This is the optimal selection of the $\DD$ matrix as presented in the main text, which significantly accelerates convergence of the bound computation.

\begin{figure}[!ht]%[p]
	\includegraphics[width=1\textwidth]{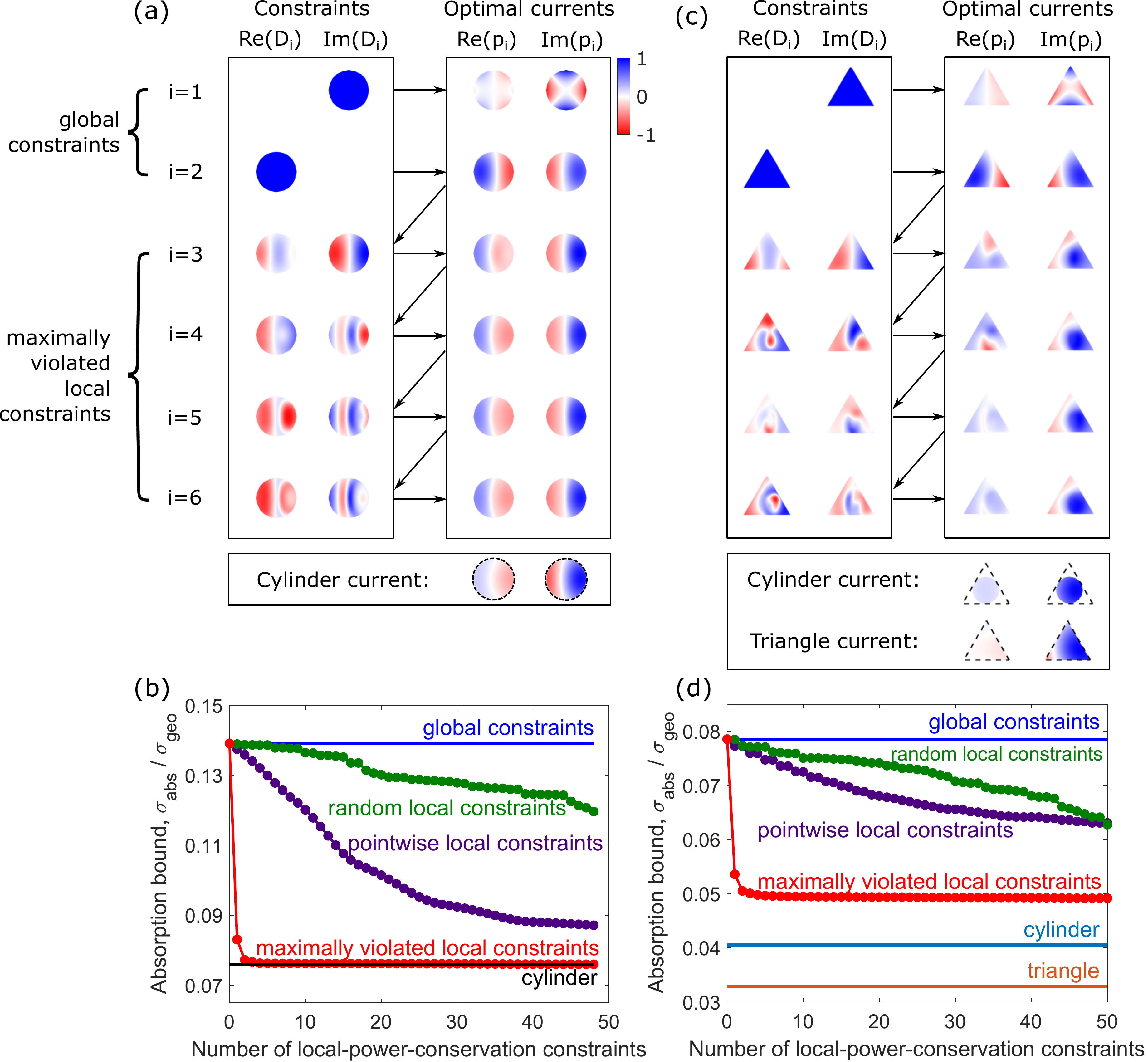}
    \centering
    \caption{\hl{(a) The left panel shows spatial profiles of the global and the first four maximally-violated local constraints. The design region is a cylinder with diameter $0.18\lambda$ and the presented profiles are its 2D cross sections. D$_i$ denotes the $E_z$ component of $\DD_i$ in space, normalized so its maximal magnitude is one. Each iteration generates an optimal current, as indicated by the direction of the arrows. The normalized $E_z$ component of the optimal currents are shown on the right panel, which in turn generate the next maximally-violated local constraints.   (b) Convergence of the upper bound on absorption cross-section, $\sigma_{\rm abs}$, under maximally-violated local constraints (red line) as compared to the ones from only the global constraints (blue line) or other type of local constraints (green and purple lines). Figures (c,d) are the same as (a,b) but the designableregion is now an equilateral triangle with side length 0.18$\lambda$.}}
    \label{fig:SM_figure1}
\end{figure}

\hl{
\section{Maximal absorption cross-section under local constraints}
In Fig. 1(b) of the main text, we provide an example of maximizing absorption cross-section under local constraints. In this section, we provide detail on the formulation of the optimization problem and the iteration process involved in identifying the maximally-violated local conservation laws. The main result in this section is summarized in \figref{SM_figure1}, where we consider not only a cylindrical design region, but also a triangular design region to showcase the generality of this computational approach.

The exact expression of absorption cross-section in terms of the polarization current $\phi$ can be identified from the global power-conservation law~\cite{Lytle2005,Newton1976,Jackson1999}, which can be derived from \eqref{loc_law} by choosing $\DD_j$ as an identity multiplied by the unit imaginary number at a real frequency $\omega$:
\begin{align}
\underbrace{\frac{\omega}{2} \phi^\dagger \left(\ImGO\right) \phi}_{\Pscat} + \underbrace{\frac{\omega}{2} \phi^\dagger \left(\Im \xi \right) \phi}_{\Pabs}  = \underbrace{\frac{\omega}{2} \Im \left(\psi_{\rm inc}^\dagger \phi\right)}_{\Pext},
\label{eq:optthm}
\end{align}
where each term from left to right represents scattered power, absorption, and extinction, respectively. If the incident wave is a plane wave, the expression of its intensity in our dimensionless unit is $I_0 = |E_0|^2 / 2$, where $E_0$ is the plane-wave amplitude. The absorption cross-section is defined as the ratio between absorption and plane-wave intensity $\sigma_{\rm abs} = P_{\rm abs} / I_0 = \omega\phi^\dagger \left(\Im \xi \right)\phi/|E_0|^2$. 

Maximizing absorption cross-section under local conservation laws is equivalent to the optimization problem:
\begin{equation}
    \begin{aligned}
        & \underset{\phi}{\text{max.}}
        & & \sigma_{\rm abs} = \omega\phi^\dagger \left(\Im \xi \right)\phi/|E_0|^2 \\
        & \text{s.t.}
        & & \phi^\dagger \Re\left\{\DD_j\omega^*(\xi+\Gamma_0)\right\} \phi = -\Re \left(\omega^*\phi^\dagger\DD_j\psiinc\right).\quad\quad j = 1, 2, ..., N
    \end{aligned}
    \label{eq:opt_prob_abs}
\end{equation}
Given designable region, incident field, and material properties as inputs, one solves this optimization problem via semidefinite relaxation discussed in Sec. \ref{sec:SDR}. The rest of this section considers a specific example where the incident wave is a TE-polarized plane wave and the material is nonmagnetic with susceptibility $\varepsilon = 12 + 0.1i$. We consider \hl{two designable regions:} a cylinder and a equilateral triangle. The cylindrical design region has diameter $d = 0.18\lambda$, total length $h$ in its translational invariant direction \hl{($h \rightarrow \infty$, so we can solve the 2D simplification)}, and a geometric cross section $\sigma_{\rm geo} = dh$. The equilateral triangle has side length $l=0.18\lambda$, total length $h$ in its translational invariant direction, and a geometric cross section $\sigma_{\rm geo} = \sqrt{3}/2lh$.

As mentioned in Sec. \ref{sec:AM}, the algorithm for generating the maximally-violated local conservation constraints is built up from the existing global conservation constraints.
Thus, we first solve the optimization problem (\ref{eq:opt_prob_abs}) with only the global real-power conservation constraint where $N = 1$ and $\DD_1 = iI$. For a cylindrical design region, the optimization program returns an optimal polarization current $p_1$ shown in the right panel of \figref{SM_figure1}(a), and an upper bound $\sigma_{\rm abs} / \sigma_{\rm geo} = 4.12$, too loose to be shown in \figref{SM_figure1}(b). Adding an additional global reactive-power conservation constraint ($N = 2$, $\DD_1 = iI$, and $\DD_2 = I$) gives us a dipole-like optimal current $p_2$ shown in \figref{SM_figure1}(a), and an upper bound $\sigma_{\rm abs} / \sigma_{\rm geo} = 0.139$, marked by the blue line in \figref{SM_figure1}(b). Next, we include extra local conservation laws in the optimization problem (\ref{eq:opt_prob_abs}) to tighten the global bound (result shown in \figref{SM_figure1}(b)). In particular, we use the algorithm derived in Sec. \ref{sec:AM} of the SM to generate maximally-violated local conservation constraints. For example, we use \eqref{D_iter} to find out a local constraint, $\DD_3$, that is maximally violated by the optimal current $p_2$. The spatial profile of its diagonal components (denoted by D$_3$) are shown in the left panel of \figref{SM_figure1}(a). This additional constraint reduces the upper bound (the second red marker from the left in \figref{SM_figure1}(b)), and together with global constraints $\DD_1$ and $\DD_2$, predicts an optimal current $p_3$ which resembles the polarization current in an unpatterned cylinder.
We continue this iteration for 50 more times in \figref{SM_figure1}(b) and show the spatial profile of the first six in \figref{SM_figure1}(a). After the fourth iteration, both the upper bound (red line \figref{SM_figure1}(b)), and the optimal currents have converged to the solution of an unstructured cylinder, suggesting the ineffectiveness of structuring in this particular case. 

The same algorithm is applied to a equilateral triangular design region shown in \figref{SM_figure1}(c,d). In this example, we consider two \hl{possible scattering} structures: an unpatterned triangle with a dimension the same as the design region, and the largest unpatterned cylinder that can fit in the design region (bottom panel of \figref{SM_figure1}(c)). Neither structure generate\hl{s} the optimal current distribution predicted in the right panel of \figref{SM_figure1}(c), and consistently, neither reach the predicted upper bound in \figref{SM_figure1}(d). Unlike a cylindrical design region where an unpatterned cylinder is already the optimum, a triangular design region may benefit from \hl{a more complex structure. From a computational perspective, the asymmetry of the triangular region has no effect on the speed or convergence of the bound computations.}
}

\section{Volume integral form of $T$-matrix}
\hl{In the main text, one of the examples considered is whether a specific scattering-matrix can be targeted by some designable region, an example that we discuss more in the next section. In this section, in preparation for that, we derive the transition-matrix ($T$-matrix) elements for waves impinging upon and exiting from a 2D circular bounding region. The $T$-matrix calculation is simpler than a direct $S$-matrix calculation, and the two are related in a simple way, as noted in the next section.}

We first derive the volume integral form of $T$-matrix elements as a linear function of the polarization current in arbitrary basis functions. Then, specifically for a 2D circular bounding region, we derive the $T$-matrix expression in the basis of vector cylindrical waves.

Given arbitrary bounding volume $V$, a set of incoming basis $\{\psi_{\rm in,n}\}$ is defined on its surface $\partial V$ through the orthogonal relation:
\begin{equation}
-\frac{1}{4}\int_{\partial V} \psi_{\rm in, i}(\xv_s)^\dagger P(\xv_s) \psi_{\rm in, j}(\xv_s) = \delta_{ij}, \quad \quad
P = 
\begin{pmatrix}
0 & \hat{\nv}\times \\
-\hat{\nv}\times & 0
\end{pmatrix},
\label{eq:orth_relation}  
\end{equation}
with $\hat{\nv}$ being the unit normal vector. When $i=j$, the right hand side of the orthogonality relation measures the power flow of state $ \psi_{\rm in, i}$ through the surface $\partial V$. (We choose the convention pointing outward for outgoing states and inward for incoming states.)
Outgoing states can be defined as the time reverse of the incoming states:
\begin{equation}
\psi_{\rm out, i}(\xv_s) = Q\psi^*_{\rm in, i}(\xv_s),  \quad \quad
Q = 
\begin{pmatrix}
I & 0 \\
0 & -I
\end{pmatrix},
\end{equation}
where the operator $Q$ flips the sign of the magnetic field, as required by time reversing.
The incident basis $\{\psi_{\rm inc, i}\}$ is defined by a linear combination of the incoming and outgoing basis:
$\psi_{\rm inc, i} = \alpha \psi_{\rm out, i} + \beta \psi_{\rm in, i}.$
Coefficients $\alpha$ and $\beta$ depend on the exact basis one choose. For example, for vector cylindrical waves, they are both $\frac{1}{2}$. 

Incident field $\psi_{\rm inc}$ can be expanded by the incident basis with coefficients $c_{\rm inc, i}$.
Similarly, scattered field $\psi_{\rm out}$ can be expanded by the outgoing basis with coefficients $c_{\rm out, i}$. These two sets of coefficients are connected by $T$-matrix.
\begin{equation}
\psi_{\rm inc} = \sum_i c_{\rm inc, i}\psi_{\rm inc, i}, \quad\quad \psi_{\rm scat} = \sum_i c_{\rm out, i}\psi_{\rm out, i},  \quad\quad 
\begin{pmatrix}
  \\
 c_{\rm out} \\
 \
\end{pmatrix}
=
\begin{pmatrix}
  \\
 T \\
 \quad\quad\quad\quad\quad
\end{pmatrix}
\begin{pmatrix}
  \\
 c_{\rm inc} \\
 \
\end{pmatrix}.
\end{equation}
Thus, the entry $T_{ij}$ measures the ratio $c_{\rm out, i} / c_{\rm inc, j}$. In other words, when the incident field $\psi_{\rm inc} = \psi_{\rm inc, i}$, $T_{ij}$ takes the value of $c_{\rm out, i}$. Using this definition, we can express $T_{ij}$ as a linear function of polarization current $\phi$ after some mathematical manipulation:
\begin{align}
T_{ij} &= -\frac{1}{4}\int_{\partial V} \psi_{\rm out, i}^\dagger(\xv_s) P(\xv_s) \psi_{\rm scat}(\xv_s) \\
 &= -\frac{1}{4\alpha}\int_{\partial V} \psi_{\rm inc, i}^\dagger(\xv_s) P(\xv_s) \psi_{\rm scat}(\xv_s), \\
 &= -\frac{1}{4\alpha}\int_{\partial V} \int_{V} \psi_{\rm inc, i}^\dagger(\xv_s)P(\xv_s) \Gamma_0(\xv_s,\xv_v)\phi(\xv_v)
\end{align}
where we used the fact that $\psi_{\rm out,i} = \frac{1}{\alpha}\psi_{\rm inc,i} - \frac{\beta}{\alpha}\psi_{\rm in}$, and the incoming and outgoing fields are orthogonal in this inner product. To further simplify this equation, we first take its transpose, and then use the properties $P^T(\xv_s) = P(\xv_s)$ and $\Gamma_0^T(\xv_s,\xv_v)=\Gamma_0(\xv_v,\xv_s)$ to write $T_{ij}$ as:
\begin{align}
T_{ij} &= -\frac{1}{4\alpha}\int_{\partial V} \int_{V} \phi^T(\xv_v)\Gamma_0(\xv_v,\xv_s)P(\xv_s)\psi_{\rm inc, i}^*(\xv_s) \\
 &= \frac{1}{4\alpha}\int_{\partial V} \int_{V} \phi^T(\xv_v)\Gamma_0(\xv_v,\xv_s)P(\xv_s)\psi_{\rm inc, i}(\xv_s),
\end{align}
where we use the properties  $\psi_{\rm inc, i}^*(\xv_s) = Q\psi_{\rm inc, i}(\xv_s)$ and $-P(\xv_s)Q= P(\xv_s)$ to derive the second equality. Lastly, we identify that the product $P(\xv_s)\psi_{\rm inc, i}(\xv_s)$ gives the surface equivalent current $\xi_{\rm inc, i}(\xv_s)$ on the surface $\partial V$, which can be propagated back to the volume through the Green's function:
\begin{align}
T_{ij} &=\frac{1}{4\alpha}\int_{\partial V} \int_{V} \phi^T(\xv_v)\Gamma_0(\xv_v,\xv_s)\xi_{\rm inc, i}(\xv_s)    \\
 &= \frac{1}{4\alpha}\int_{V} \phi^T(\xv_v)\psi_{\rm inc, i}(\xv_v) \\
 &= \frac{1}{4\alpha}\phi^T\psi_{\rm inc, i}. \label{eq:Tij0}
\end{align}
The key result, \eqref{Tij0}, identifies $T_{ij}$ as a overlap integral between incident channel $\psi_{\rm inc, i}$ and polarization current $\phi$ that is induced by incident field $\psi_{\rm inc, j}$.

For a highly symmetric bounding volume, the derivation of the volume integral form of $T$-matrix can be greatly simplified. In the example provided in the main text, we assume nonmagnetic material with a 2D bounding area and TE incidence. The basis for outgoing and incident field can be chosen as the set of vector cylindrical waves:
\begin{align}
v_{\rm inc, n}(\xv) &= \frac{1}{2}\hat{z}J_n(k\rho)e^{in\phi} \label{eq:vcw_1} \\
v_{\rm out, n}(\xv) &= \frac{1}{2}\hat{z}H^{(1)}_n(k\rho)e^{in\phi}, \label{eq:vcw_2}
\end{align}
where $J_n(x)$ is the Bessel function of order $n$, and $H^{(1)}_n(x)$ is the Hankel function of the first kind of order $n$. Conventionally, these two basis written here do not include magnetic field, and are not normalized based on \eqref{orth_relation}, so we use different notations other than $\psi_{\rm inc, n}$ and $\psi_{\rm out, n}$.

As discussed before, when $\psi_{\rm inc} = v_{\rm out, j}$, the entry $T_{ij}=c_{\text{scat}, i}$. By virtue of the Green's function expansion $\GO(\xv, \xv') = i \sum_n v_{\rm out, n}(\xv) v_{\rm inc, n}^\dagger(\xv')$ for $\rho>\rho'$, we can easily derive the volume integral form of $T_{ij}$:
\begin{align} 
	T_{ij}	&= \frac{\int_{\partial V} v^\dagger_{\rm out, i}(\xv_s)\psi_{\rm scat}(\xv_s)\td S}{\int_{\partial V} v^\dagger_{\rm out, i}(\xv_s)v_{\rm out, i}(\xv_s)\td S} \\
			&= \frac{\int_{\partial V} v^\dagger_{\rm out, i}(\xv_s)  \left[\int_V \GO(\xv, \xv')\phi(\xv') \td V\right]     \td S}{\int_{\partial V} v^\dagger_{\rm out, i}(\xv_s)v_{\rm out, i}(\xv_s)\td S} \\
			&= \frac{\int_{\partial V} v^\dagger_{\rm out, i}(\xv_s)  \left[\int_V i\left(\sum_n v_{\rm out, n}(\xv) v_{\rm inc, n}^\dagger(\xv') \right)\phi(\xv) \td V\right]     \td S}{\int_{\partial V} v^\dagger_{\rm out, i}(\xv_s)v_{\rm out, i}(\xv_s)\td S} \\
			&= i\int_V v_{\rm inc, i}^\dagger(\xv')\phi(\xv')\td V \\
			&= i v_{\rm inc, i}^\dagger \phi
\end{align}
Similar as before, the result suggests that $T_{ij}$ is the projection of $\phi$ into the given incident basis $v_{\rm inc, i}$ with an additional phase delay, under the incident field $\psi_{\rm inc} = v_{\rm inc, j}$. There is slight difference between this and the more general result in \eqref{Tij0} because the vector cylindrical waves defined in  equations (\ref{eq:vcw_1}) and (\ref{eq:vcw_2}) do not include magnetic field components and are not normalized based on \eqref{orth_relation}.

\section{Formulation of $S$-matrix feasibility bound}
The objective for the S-matrix feasibility problem is to minimize the relative difference between the achievable and target $S$ matrices:
\begin{equation}
\text{Min\ \ } f_{\rm obj} = \left\| S - \St \right\|^2 / \left\|\St\right\|^2,
\label{eq:fobj}
\end{equation}
where we choose $\|\cdot\|$ to denote Frobenius norm. 

It is simpler to translate the scattering matrix $S$, which relates \emph{incoming} waves to \emph{outgoing} waves, into the transition matrix $T$, which relates \emph{incident} waves to \emph{scattered} waves. One can typically choose a basis \hl{(such as the cylindrical-wave basis) for which} $S = I+2T$. Inserting this relation into \eqref{fobj}, we have:
\begin{align}
f_{\rm obj} &= 4\left\| T - \Tt \right\| / \left\|\St\right\|^2 \\
	 &= \frac{4}{||\St||^2}\sum_{ij}|T_{ij} - T_{\text{target}, ij}|^2 \\
	 &= \frac{4}{||\St||^2}\sum_j f_{\text{obj}, j},
\end{align}
where in the last equality we separate out the objective into contributions from different incident fields:
\begin{align}
	 f_{\rm obj,j} = \sum_i |T_{ij} - T_{\text{target}, ij}|^2. \label{eq:fobj_j} 
\end{align}	
Each $f_{\rm obj,j}$ corresponds to the scattering from incident field indexed by $j$, so we bound them separately and later add up their contributions. As we proved in section IV of the SM, $T_{ij}$ can be written as a linear function of $\phi$, which is the induced polarization current under the incident field $\psi_{\rm inc} = \psi_{\rm inc, j}$. Assume this linear relation is $T_{ij}=w_i^\dagger \phi$. We can plug it in \eqref{fobj_j} to express each $f_{\rm obj,j}$ as a quadratic function of $\phi$:
\begin{align}
	 f_{\rm obj,j} = \phi^\dagger\left(\sum_iw_i w_i^\dagger\right)\phi + \Re\left[\left(-2\sum_i T_{\text{target}, ij}w_i\right)^\dagger\phi\right] + \sum_i|T_{\text{target}, ij}|^2, \label{eq:Sobj}
\end{align}	
This can be written in the form of optimization problem (3) in the main text (after adding a minus sign to the objective to turn minimization into maximization) with $\mathbb{A} = -\sum_iw_iw_i^\dagger$, $\beta = 2\sum_i T_{\text{target}, ij}w_i$, $c = -\sum_i|T_{\text{target}, ij}|^2$, and $\psi_{\rm inc} = \psi_{\rm inc, j}$. 

For the general case where the incident basis $\psi_{\rm inc, i}$ is defined through \eqref{orth_relation}, we substitute $\omega_i$ in \eqref{Sobj} with $\psi_{\rm inc, i}^* / (4\alpha^*)$. For the specific case where the we assume nonmagnetic material with a 2D bounding area and TE incidence, the incident basis $\psi_{\rm inc, i}$ is vector cylindrical waves $v_{\rm inc, i}$ defined in \eqref{vcw_1}, and we substitute $\omega_i$ with $-iv_{\rm inc, i}$.

%\section{Single-frequency extinction bound and example of $\DD$ matrices}
%figure below

%\begin{figure}[p]
%	\includegraphics[width=0.5\textwidth]{SM_figure3}
%    \centering
%    \caption{Bounds on the single-frequency extinction coefficient, $Q_{\rm ext}$, as a function of frequency, $\omega$, for a Drude-Lorentz material with plasmonic frequency $\omega_p$, loss rate $\gamma=0$, and resonant frequency $\omega_0/\omega_p = 0.3015$, and is enclosed by a bounding area with $R=1.5/\omega_p$.
%    Gray shading marks the computationally difficult region, where the permittivity $\varepsilon$ is too large for the program to numerically discretize the structure. Two local power conservation laws, in addition to the global ones, are leveraged for the red dotted line.}
%    \label{fig:SM_figure3}
%\end{figure}

%\section{Local linear and angular momentum conservation constraint}

\section{Minimum diameter of a power splitter}
\begin{figure}[!ht]%[p]
	\includegraphics[width=1\textwidth]{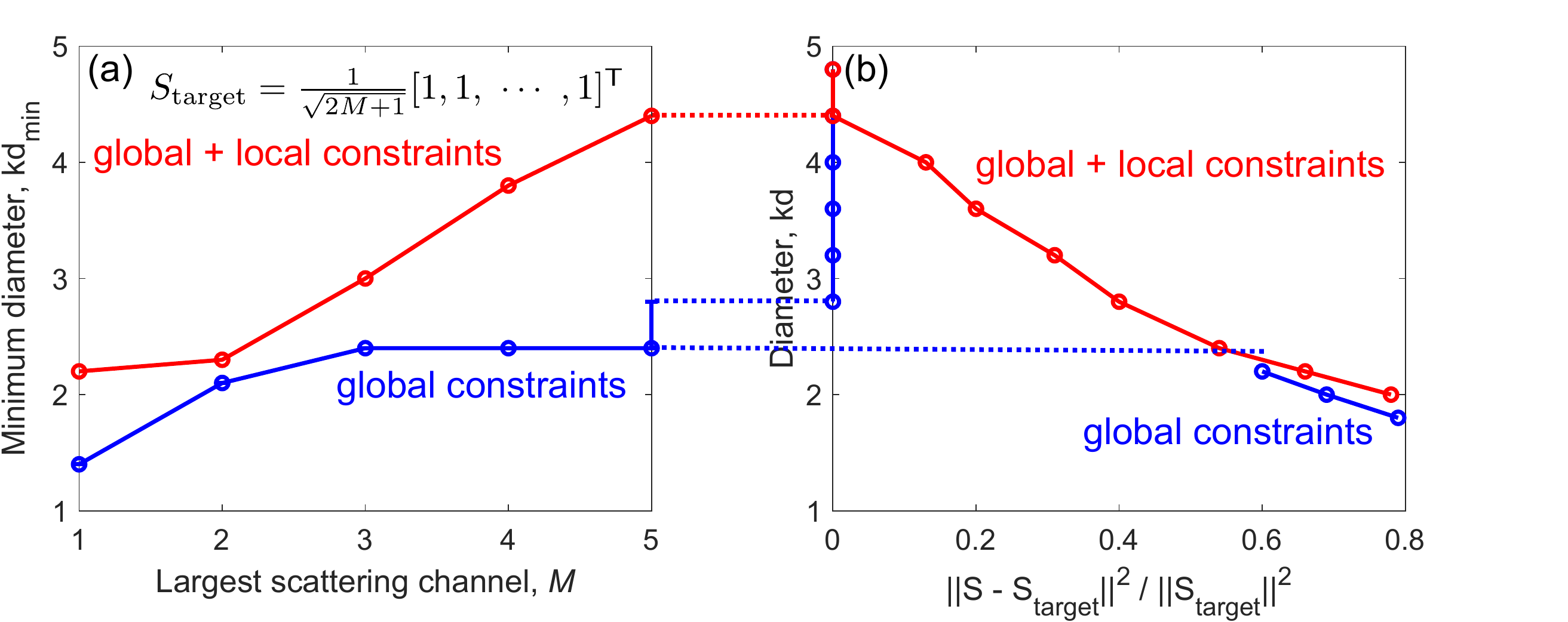}
    \centering
    \caption{(a) Minimum diameter required for a power splitter for a single input to 2M + 1 outgoing channels. (b) Lower bounds on the objective $||S-S_{\rm target}||^2 / ||\St||^2$ at each diameter when the largest scattering channel $M=5$. Uncertainties rising from the numerical instabilities in global constraints are marked by the dotted blue lines.}
    \label{fig:SM_figure4}
\end{figure}

In the main text, we show the minimum diameter required for a power splitter for a single input to 2M + 1 outgoing channels in the cylindrical-wave basis. The way we determine the minimum diameter for each $M$ is to minimize the objective function $\left\| S - \St \right\|^2 / \left\|\St\right\|^2$ for every diameter $d$, and choose the smallest one that satisfies $\left\| S - \St \right\|^2 / \left\|\St\right\|^2 < 1\%$. This process is shown in \figref{SM_figure4}(b) for the case with $M=5$.

The gap between two blue lines in \figref{SM_figure4}(b) originates from a numerical instability in the global-constraint-only approach. Higher orders of the cylindrical waves yield widely separated numerical scales in the corresponding matrices, such that with only global constraints the optimization does not terminate successfully for some diameters. The two dashed blue lines indicate the uncertainty region for determining the minimum diameter. The lower bound of this uncertainty region is estimated from the asymptotic limit of the global-constraint-only approach in \figref{SM_figure4}(a). The minimum diameter can be lower bounded by the lower dashed line of the uncertainty region, which explains the location of the circular point with the errorbar.

\section{Formulation of the bandwidth-averaged extinction bound}
In this section, we transform the bandwidth-averaged extinction to a single scattering amplitude at a complex frequency by Cauchy's residue theorem, using a similar technique to that which has been demonstrated in Refs.~\cite{Hashemi2012,Shim2019}.
We start with the expression of single-frequency extinction cross section at a real frequency:
\begin{align}
	\sigma_{\rm ext}(\omega) = \Im\left[\omega \psi^\dagger_{\rm inc}(\omega) \phi(\omega) \right] \label{eq:ecs-1}
\end{align}
Incident field $\psi_{\rm inc}(\omega)$ in far-field scattering is often approximated as a plane wave. Without loss of generality, we assume it has unit intensity and is propagating along the $x$ direction. We use dimensionless quantities with $c=1$, so the plane-wave frequency dependence can be written as $e^{i\omega x}$. In anticipation of an analytic continuation into the complex plane, we use the general relation $\psi_{\rm inc}^*(\omega) = \psi_{\rm inc}(-\omega)$ for real-valued frequencies~\cite{Landau1960} to remove the complex conjugation (which cannot be analytically continued):
\begin{align}
	\sigma_{\rm ext}(\omega) &= \Im\left[\omega \psi^T_{\rm inc}(-\omega) \phi(\omega) \right] \label{eq:ecs-2} \\
							&= \Im s(\omega). \label{eq:ecs-3}
\end{align}
Here, we define a new term $s(\omega)=\omega \psi^T_{\rm inc}(-\omega) \phi(\omega)$ that we identify as the far-field scattering amplitude.
Since the incident plane wave $\psi_{\rm inc}(\omega)$ has the frequency dependence $e^{i\omega x}$ (analytic everywhere), and the polarization current $\phi(\omega)$ is a causal linear-response function~\cite{Nussenzveig1972}, the amplitude $s(\omega)$ is analytic in the upper half of the complex-frequency plane (UHP).

The average extinction cross section $\langle \sigma_{\rm ext}\rangle$ in a bandwidth $\Delta\omega$  around a center frequency $\omega_0$ can be defined as the integral of the product of $\sigma_{\rm ext}(\omega)$ and a Lorentzian window function $\Hw(\omega)=\frac{\Delta\omega/\pi}{(\omega-\omega_0)^2+\Delta\omega^2}$: 
\begin{align}
	\langle \sigma_{\rm ext}\rangle &= \int_{-\infty}^{+\infty} \sigma_{\rm ext}(\omega)H(\omega)\td\omega  \label{eq:int_1}\\
	&= \Im\int_{-\infty}^{+\infty} s(\omega)H(\omega)\td\omega \label{eq:int_2}  % \\
	% &= 	\Im\left[\tomega \psi^T_{\rm inc}(-\tomega)\phi(\tomega)\right] 
\end{align}
The integrand $s(\omega)H(\omega)$ has two properties that allows us to use Cauchy’s residue theorem to equate
the all-frequency integral to a single pole in the UHP. The first property is that the $s(\omega)H(\omega)$ only has one pole $\tomega = \omega_0 + i\Delta\omega$ from the window function in the UHP, since $s(\omega)$ is complex analytic in the UHP as discussed above. The second property is the magnitude of  $s(\omega)H(\omega)$ decays faster than $1/|\omega|$ when $|\omega|\rightarrow +\infty$. In this asymptotic limit, the window function $H(\omega)$ decays at a rate of $1/|\omega|$, and the amplitude $s(\omega)$ decay at the rate of $1 / |\omega|$, which can be proved as follows.

In the high-frequency limit, the polarization field must decay towards zero (the bound charges cannot respond to
such high frequencies), and on physical grounds~\cite{Landau1960} the decay must occur in proportion to $1/|\omega|$. Conventionally, the decay constant is chosen to be a ``plasma frequency'' $\omega_p$ that is physically meaningful for metals but applies to dielectrics as well. Because the scatterer becomes transparent at high frequencies, the Born approximation applies and the polarization field will be directly proportional to the incident field: $\phi(|\omega|\rightarrow\infty) = -\frac{\omega_p^2}{\omega^2}\psi_{\rm inc}(\omega)$, so that $s(|\omega|\rightarrow\infty) = -\frac{\omega_p^2}{\omega}\psi_{\rm inc}^T(-\omega)\psi_{\rm inc}(\omega) \sim 1/\omega$. Note that the inner product $\psi_{\rm inc}^T(-\omega)\psi_{\rm inc}(\omega)$ does not dependent on frequency as the frequency dependence of the incident plane wave is just $e^{i\omega x}$.

Taking these two properties into account, we can connect the upper and lower limit of the integral in \eqref{int_1} by a half circle in the UHP, which does not actually contribute to the integral due to the fast decay rate of the $s(\omega)H(\omega)$. Integration of this closed loop can be transformed into the single pole of $s(\omega)H(\omega)$ at $\tomega = \omega_0 + i\Delta\omega$ by Cauchy’s residue theorem, giving the expression in the main text:
\begin{equation}
\langle \sigma_{\rm ext}\rangle = \Im\left[\tomega \psi^T_{\rm inc}(-\tomega)\phi(\tomega)\right].
\end{equation}
In the case of TE incidence in a 2D geometry with nonmagnetic material, we only need to consider the $z$ polarization component of the electric incident field, which is a scalar quantity. If we still use notation $\psi_{\rm inc}$ to denote this quantity, we can solve for the maximum $\langle \sigma_{\rm ext}\rangle$ by the optimization problem with $\beta = i\tomega^*e^{i\omega_0x+\Delta\omega x}$ and incident field $\psi_{\rm inc}(\tomega)=e^{i\omega_0x-\Delta\omega x}$.

\section{Positive semidefinite property of scattering and absorption operators}
The power-bandwidth limit discussed in the main text relies on the fact that the local energy conservation laws can be extended to complex frequency $\omega$. Explicitly writing out the frequency dependency of the operators, and introducing $\xi(\omega)=-\chi(\omega)^{-1}$, we have:
\begin{align}
\frac{\omega^*}{2} \phi^\dagger \DD\GO(\omega) \phi + \frac{\omega^*}{2} \phi^\dagger \DD\xi(\omega) \phi  = -\frac{\omega^*}{2} \phi^\dagger\DD\psi_{\rm inc}.
\label{eq:optthm2}
\end{align} 
Among all the possible local conservation laws we can impose, the most important one is the global power-conservation law. It constrains the optimization variable $\phi$ to the boundary of a high-dimensional ellipsoid, and can be derived by assigning $\DD$ an identity tensor and take the imaginary part of \eqref{optthm2}: 
\begin{align}
\frac{1}{2} \phi^\dagger \Im\left[\omega^*\GO(\omega)\right] \phi + \frac{1}{2} \phi^\dagger \Im\left[\omega^*\xi(\omega)\right] \phi  = -\frac{1}{2} \Im(\omega^*\phi^\dagger\psi_{\rm inc}).
\label{eq:global_power}
\end{align} 
In this section, we prove the positive semidefinite property of the two involving operators, $\Im\{\omega^*\GO(\omega)\}$ and $\Im\{\omega^*\xi(\omega)\}$, in the UHP, using a similar technique to that which has been used in Ref.~\cite{zemanian_hilbert_1970, zemanian_realizability_1995, welters_speed--light_2014}.

We first prove the positive semidefinite property of the operator $\Im\{\omega^*\xi(\omega)\}$ in a passive scattering problem. Passivity requires that the polarization currents $\phi$ in the material do not do work. The total work they do up to a time $t$ must be greater than or equal to zero:
\begin{equation}
    \Re\int {\rm d}x \, \int_{-\infty}^{t} {\rm d}t'\, \psi^\dagger(x,t') \frac{\td\phi(x,t')}{\td t'} \geq 0.
\label{eq:pvt1}
\end{equation}
In a scattering problem where $\psi$ is the total field, we can interpret $\phi$ as the polarization currents, which are the convolution of the susceptibility in time and space (we allow for spatial nonlocality):
\begin{align}
    \phi(x,t') = \int {\rm d}x' \int_{-\infty}^{t'} {\rm d}t'' \, \chi(x,x',t'-t'') \psi(x',t'') = \int {\rm d}x' \int_0^\infty \chi(x,x',\tau) \psi(x',t'-\tau) \,{\rm d}\tau,
\end{align}
where in the second expression the variable $\tau$ can be interpreted as the delay since the excitation that is creating a response. Inserting the latter expression into \eqref{pvt1} we have:
\begin{align}
    \Re\int \int {\rm d}x' {\rm d}x \, \int_{-\infty}^{t} {\rm d}t'\, \psi^\dagger(x,t') \int_0^{\infty} {\rm d}\tau \, \chi(x,x',\tau) \psi'(x',t'-\tau) \geq 0,
    \label{eq:TI1}
\end{align}
where $\psi'$ denotes the derivative of $\psi$. The expression of \eqref{TI1} must be valid for all $\psi$. We can choose a simple time-dependence for $\psi$, following Refs.~\cite{zemanian_realizability_1995, welters_speed--light_2014}:
\begin{align}
    \psi(x,t') = 
    \begin{cases}
        \psi(x) e^{-i\omega t'} & \text{ for } t' < T \\
        0 & \text{ for } t' \geq T,
    \end{cases}
\end{align}
where $\omega$ is a complex-valued frequency, i.e. $\omega = \omega_0 + i \Im \omega$, and $T$ is simply a shut-off time that we will always choose larger than $t$ and which assures technical conditions are satisfied in rigorous proofs~\cite{zemanian_realizability_1995, welters_speed--light_2014}. Given this form, \eqref{TI1} becomes:
\begin{align}
    \Re\int \int {\rm d}x' {\rm d}x \, \psi^{\dagger}(x) \int_{-\infty}^{t} {\rm d}t'\, e^{i\omega_0 t'} e^{(\Im \omega) t'} \int_0^{\infty} {\rm d}\tau \, \chi(x,x',\tau) \left(-i\omega\right) e^{-i\omega_0 (t'-\tau)} e^{(\Im \omega) (t'-\tau)} \psi(x') \geq 0,
    \label{eq:TI2}
\end{align}
Re-arranging terms then gives
\begin{align}
    \Re \left[ \left(-i\omega\right) \int \int {\rm d} x' {\rm d}x \, \psi^{\dagger}(x) \int_{-\infty}^{t} {\rm d}t'\, e^{2(\Im \omega) t'} \left\{ \int_0^{\infty} {\rm d}\tau \, \chi(x,x',\tau) e^{i\omega \tau} \right\} \psi(x') \right] \geq 0.
    \label{eq:TI3}
\end{align}
The term in curly brackets is proportional to the Fourier transform of $\chi$, i.e. $\chi(\omega)$ at complex frequency $\omega$, and we can drop the constants related to $2\pi$. The integral over $t'$ is easily evaluated. Finally, noting that $\Re(-iz) = \Im(z)$, we have the expression
\begin{align}
    \frac{e^{2(\Im\omega)t}}{2 \Im \omega} \Im \int \int {\rm d} x' {\rm d}x \, \psi^{\dagger}(x) \left[ \omega \chi(x,x',\omega) \right] \psi(x') \geq 0.
    \label{eq:TI4}
\end{align}
This expression must be valid for all $\psi(x)$ distributions. We can remove the spatial dependence of $\chi$ and instead treat it as a square matrix (as in any standard discretization), in which case we can simply write that
\begin{align}
    \Im \left[ \omega \chi(\omega) \right] \geq 0 \quad \text{ for } \Im \omega > 0,
    \label{eq:chipos}
\end{align}
where the imaginary part of the matrix argument refers to its anti-Hermitian part; e.g., $\Im A = (A - A^\dagger)/2i$. 

To convert \eqref{chipos} to an inequality for $\xi$, we use the fact that $\chi = -\xi^{-1}$ to rewrite \eqref{chipos} as
\begin{align}
    \Im \left[ \omega \chi \right] &= \Im \left[-\omega \xi^{-1} \right] \\
                                   &= \Im \left[-\omega \left(\xi^\dagger\xi\right)^{-1} \xi^\dagger\right] \\
                                   &= \left(\xi^\dagger\xi\right)^{-1} \left( \Im \left[ \omega^* \xi \right] \right),
\end{align}
which implies that
\begin{empheq}[box=\widefbox]{align}
    \Im \left[ \omega^* \xi(\omega) \right] \geq 0 \quad \text{ for } \Im \omega > 0.
\end{empheq}

Thus we have our proof for the positive semidefinite property of the first of our two operators. Now we can follow similar logic for the second one. We start with an expression similar to \eqref{pvt1}, but now we change our interpretation: we will take the $\phi$ to be free currents, $\psi$ to be the fields radiated by them, and the quantity in \eqref{pvt1} then represents the \emph{negative} of the work done by those currents on the outgoing field (which again must be positive). Thus our starting point is the negative of \eqref{pvt1}:
\begin{equation}
    \Re\left[ - \int {\rm d}x \, \int_{-\infty}^{t} {\rm d}t'\, \frac{\td\phi(x,t')}{\td t'}^\dagger \psi(x,t')  \right] \geq 0,
\label{eq:pvt2}
\end{equation}
where we also reversed the order of our arguments in the integrand for simplicity below. (That is allowed because $\Re z = \Re z^*$.)

Now our convolution relation will connect the fields at a time $t'$ to the polarization currents at an earlier time through the background Green's function $\Gamma$:
\begin{align}
    \psi(x,t') = \int {\rm d}x' \int_{-\infty}^{t'} {\rm d}t'' \, \Gamma(x,x',t'-t'') \phi(x',t'') = \int {\rm d}x' \int_0^\infty \Gamma(x,x',\tau) \phi(x',t'-\tau) \,{\rm d}\tau.
\end{align}
We are going to insert this convolution relation into \eqref{pvt2}, analogous to what we did before. We can combine this with the step of specifying a time-dependence for the function $\phi(x,t)$:
\begin{align}
    \phi(x,t') = 
    \begin{cases}
        \phi(x) e^{-i\omega t'} & \text{ for } t' < T \\
        0 & \text{ for } t' \geq T.
    \end{cases}
\end{align}
Performing these two steps in \eqref{pvt2} we have:
\begin{equation}
    \Re\int\int {\rm d}x {\rm d}x' \, \int_{-\infty}^{t} {\rm d}t'\, (-i\omega^*) e^{i\omega_0 t'} e^{(\Im \omega)t'} \phi^\dagger(x) \int {\rm d}\tau \, \Gamma(x,x',\tau) e^{-i\omega(t'-\tau)} \phi(x') \geq 0.
\label{eq:TI1b}
\end{equation}
As before, the oscillatory terms cancel, the integral over $\tau$ is proportional to the $\Gamma(x,x',\omega)$, i.e. the Fourier transform of $\Gamma(x,x',\tau)$, and the integral over $t'$ is simple to do. We are left with:
\begin{align}
    \frac{e^{2(\Im\omega)t}}{2 \Im \omega} \Im \int \int {\rm d}x {\rm d}x' \, \phi^{\dagger}(x) \left[ \omega^* \Gamma(x,x',\omega) \right] \phi(x') \geq 0.
    \label{eq:TI4b}
\end{align}
If we again treat $\Gamma$ in space as a square matrix, we thus have
\begin{empheq}[box=\widefbox]{align}
    \Im \left[ \omega^* \Gamma(\omega) \right] \geq 0 \quad \text{ for } \Im \omega > 0,
\end{empheq}
where again the imaginary part of the matrix refers to its anti-Hermitian part.

\bibliography{lc_bib}